\documentstyle{mn}
\input{psfig.sty}
\newcommand{\etal}{{et al.~}}
\newcommand{\lta}{\la}
\newcommand{\gta}{\ga}

\newcommand{\kmsmpc}{\>{\rm km}\,{\rm s}^{-1}\,{\rm Mpc}^{-1}}
\newcommand{\kms}{\>{\rm km}\,{\rm s}^{-1}}

\newcommand{\Mpc}{\>{\rm Mpc}}

\newcommand{\Msun}{\>{\rm M_{\odot}}}
\newcommand{\msun}{\>{\rm M_{\odot}}}
\newcommand{\Lsun}{\>{\rm L_{\odot}}}
\newcommand{\MLsun}{\>({\rm M}/{\rm L})_{\odot}}

\newcommand{\reference}{\bibitem}
\newcommand{\beq}{\begin{equation}}
\newcommand{\eeq}{\end{equation}}

\newcommand{\mpch}{\>h^{-1}{\rm {Mpc}}}

\newcommand{\msunh}{\>h^{-1}\rm M_\odot}

\newcommand{\walpha}{\tilde{\alpha}}
\newcommand{\wLstar}{\tilde{L}^{*}}
\newcommand{\hxi}{\hat{\xi}}

\newcommand{\apj}{ApJ}
\newcommand{\apjs}{ApJS}
\newcommand{\aj}{AJ}
\newcommand{\mnras}{MNRAS}
\newcommand{\aap}{A\&A}

\newcommand{\pasp}{PASP}

\newdimen\hssize
\newdimen\hdsize 
\begin{document}


\title[The Conditional Luminosity Function]
      {Constraining Galaxy Formation and Cosmology with
       the Conditional Luminosity Function of Galaxies}
\author[Yang, Mo \& van den Bosch]
       {Xiaohu Yang$^{1,2}$, H.J. Mo$^{1}$ and Frank C. van den Bosch$^{1}$ 
        \thanks{E-mail: xhyang@mpa-garching.mpg.de}\\
        $^1$Max-Planck-Institut f\"ur Astrophysik, Karl Schwarzschild
         Str. 1, Postfach 1317, 85741 Garching, Germany\\
        $^2$Center for Astrophysics, University of Science and Technology
         of China, Hefei, Anhui 230026, China}


\date{Accepted ........
      Received .......;
      in original form .......}


\maketitle

\label{firstpage}


\begin{abstract}
  We  use the conditional luminosity   function  $\Phi(L \vert  M){\rm
  d}L$,  which gives the number  of galaxies  with luminosities in the
  range $L \pm {\rm d}L/2$ that reside in a halo  of mass $M$, to link
  the  distribution   of  galaxies  to that    of  dark matter haloes.
  Starting from the number density  of dark matter haloes predicted by
  current models of structure  formation, we seek  the form of $\Phi(L
  \vert   M)$ that reproduces the  galaxy  luminosity function and the
  luminosity dependence of the  galaxy  clustering strength.  We  test
  the models   of $\Phi(L \vert     M)$  by comparing  the   resulting
  mass-to-light ratios  with constraints  from  the Tully-Fisher  (TF)
  relation and   from  galaxy clusters.   A  comparison  between model
  predictions and current observations   yields a number of  stringent
  constraints on both  galaxy formation and cosmology.  In particular,
  this method can   break the degeneracy  between $\Omega_0$   and the
  power-spectrum   normalization    $\sigma_8$,   inherent  in current
  weak-lensing and  cluster-abundance  studies.  For flat $\Lambda$CDM
  cosmogonies with  $\sigma_8$ normalized by recent weak gravitational
  lensing observations, the best  results  are obtained for  $\Omega_0
  \sim 0.3$; $\Omega_0 \lta 0.2$ leads to too large galaxy correlation
  lengths,  while $\Omega_0  \gta  0.4$  gives too high  mass-to-light
  ratios to  match the observed TF relation.   The best-fit  model for
  the   $\Lambda$CDM  concordance  cosmology  with  $\Omega_0=0.3$ and
  $\Omega_{\Lambda}=0.7$   predicts   mass-to-light  ratios   that are
  slightly too high to match the TF  relation.  We discuss a number of
  possible effects that  might   remedy this problem, such    as small
  modifications of $\sigma_8$ and the Hubble parameter with respect to
  the  concordance  values,  the  assumption  that  the   universe  is
  dominated  by  warm    dark matter, systematic   errors   in current
  observational data, and the existence of dark  galaxies.  We use the
  conditional luminosity  function  derived from the present   data to
  predict several statistics about the distribution of galaxy light in
  the local Universe.  We show that  roughly $50$ percent of all light
  is  produced in haloes less massive  than $2 \times 10^{12} \msunh$.
  We also derive the probability distribution  $P(M \vert L) {\rm d}M$
  that a galaxy of luminosity $L$ resides in a halo with virial masses
  in the range $M\pm {\rm d}M/2$.
\end{abstract}




\section{Introduction}
\label{sec:intro}

It is  well established that  galaxies reside in extended  dark matter
haloes.   Because of  this, the  standard assumptions  in  the current
paradigm  of structure formation  are that  a weakly  interacting mass
component (such as the cold  dark matter, hereafter CDM) dominates the
mass  of  the Universe  and  that galaxies  form  by  the cooling  and
condensation of gas in  dark matter haloes.  Numerical simulations and
analytical models  can now  give us a  fairly detailed picture  of the
abundance  and clustering  of CDM  haloes  (e.g.  Mo  \& White  2002).
However, in order to build  a coherent picture of galaxy formation, we
need to  be able to  link these properties  of the halo  population to
those of  the galaxy population. In  other words, we need  to know how
dark matter haloes are populated by galaxies.

Three different approaches have been used in the past to link galaxies
to dark matter haloes.  The first method tries to infer the properties
of a  halo from the properties  of its galaxy  population.  The second
method relies on {\it ab initio} models for the formation of galaxies,
and the third method attempts to link the distributions of dark matter
haloes and galaxies using a more statistical approach.

Mass estimates of  individual dark matter haloes  can be obtained from
galaxy kinematics,  from gravitational lensing,  or from  studying the
X-ray  haloes.    However,  each  of  these    methods  has its    own
shortcomings: kinematics generally only   probe the inner  regions  of
dark matter haloes, results from  gravitational lensing depend  rather
sensitively on  the models adopted for the  lensing systems, and X-ray
measurements have to   rely  on the assumption  that  the  gas   is in
hydrostatic equilibrium.  Although significant  progress has been made
in each of these areas, the main limitation is  that all these methods
yield mass-to-light ratios  for individual  objects only.  Because  of
this, various authors have attempted a more global approach by using a
combination   of  galaxy luminosity  functions and luminosity-velocity
relations (such  as    the Tully-Fisher   relation) to   construct   a
distribution function for  halo circular velocities (or masses), which
in   principle can be compared directly   to the results from $N$-body
simulations (i.e.,  Shimasaku  1993; Kauffmann \etal 1993;   Newman \&
Davis 2000; Gonzales \etal 2000; Bullock  \etal 2001a; Kochanek 2001).
Such  analyses  can   be used  to   put  constraints  on the   typical
mass-to-light   ratio  for a population of    galaxies, such as spiral
galaxies.    However, this method needs  to   make a rather unreliable
conversion from an observed    velocity measure, such as   the maximum
rotation velocity of a disk galaxy, to the virial velocity of the dark
matter halo (see discussions   in Gonzales \etal 2000; Kochanek  2001;
van den Bosch 2002).

An alternative  approach is  to build {\it  ab initio} models  for the
formation of galaxies, using either large numerical simulations (e.g.,
Katz,  Weinberg \&  Hernquist 1996;  Fardal \etal  2001;  Pearce \etal
2000; Kay \etal 2001), (semi-)  analytical models (e.g., White \& Rees
1978;  White  \& Frenk  1991;  Kauffmann,  White  \& Guiderdoni  1993;
Somerville \&  Primack 1999; Cole  \etal 2000; Benson \etal  2002; van
den Bosch 2001, 2002), or  a combination of both (Kauffmann, Nusser \&
Steinmetz  1997; Kauffmann  \etal  1999; Diaferio  \etal 1999;  Benson
\etal 2000,  2001, Springel \etal 2001; Mathis  \etal 2001).  Although
these  methods have  yielded many  useful predictions  about  how dark
matter  haloes  are  populated  by  galaxies  of  different  intrinsic
properties, many of the physical processes (such as star formation and
feedback) involved are  still poorly understood at the  present and so
some uncertain  assumptions have  to be made  about a number  of model
ingredients such as the efficiency of star formation and feedback, the
stellar initial mass function, and the amount of dust extinction.

With the advent of large, homogeneous data sets such as the Two Micron
All Sky  Survey (2MASS), the  Two Degree Field Galaxy  Redshift Survey
(2dFGRS), and the Sloan Digital Sky Survey (SDSS), it becomes possible
to address  the link between dark  matter haloes and  galaxies using a
more statistical  approach.  The population of  galaxies in individual
haloes  can formally  be  represented by  a halo-occupation  function,
$P(N\vert M)$,  which gives  the probability that  a halo of  mass $M$
contains  $N$ galaxies (with  certain intrinsic  properties). Together
with  an assumption  about  the spatial  distribution  of galaxies  in
individual  haloes,  $P(N\vert  M)$  is sufficient  to  transform  the
abundance  and clustering  of the  halo population  into those  of the
galaxy population.  Numerous  studies in the past have  used models of
$P(N  \vert  M)$  to  study  the bias  and  various  other  clustering
statistics of  the galaxy distribution  (e.g., Jing, Mo  \& B\"{o}rner
1998;  Peacock \&  Smith 2000;  Seljak 2000;  Scoccimarro  \etal 2001;
White 2001;  Jing, B\"orner  \& Suto 2002;  Berlind \&  Weinberg 2002;
Bullock, Wechsler \& Somerville  2002; Scranton 2002; Kang \etal 2002;
Marinoni \& Hudson 2002).  The  disadvantage of this method is that it
does  not  provide a  physical  understanding  of  the resulting  halo
occupation  function,   and  as  such  it  is   complementary  to  the
semi-analytical modelling described above.

Clearly, $P(N  \vert M)$ is  closely  related to how galaxies  form in
dark matter haloes, and  so is an important quantity  in the theory of
galaxy  formation.  However, $P(N  \vert M)$ only contains information
about the {\it number} of  galaxies per halo  (brighter than a certain
lower limit).  Probably   the the most  important  characteristic of a
galaxy is  its luminosity $L$.  In  this paper we therefore extend the
halo-occupation modeling  technique by   labelling the  galaxies  with
luminosities.  We use  the `conditional luminosity function',  $\Phi(L
\vert M) {\rm  d}L$, which gives the  average number  of galaxies with
luminosities in the range $L\pm {\rm d}L/2$ as a function of halo mass
$M$, to address the clustering and abundance of  galaxies {\it both as
function  of luminosity}.  It   is easy to   see that the  conditional
luminosity    function provides a   direct    link between  the galaxy
luminosity function (hereafter LF) $\Phi(L) {\rm d}L$, which gives the
number of galaxies per comoving volume  with luminosities in the range
$L\pm {\rm d}L/2$, and the  halo mass function  $n(M) {\rm d}M$, which
gives the number of dark matter haloes per comoving volume with masses
in the range $M \pm {\rm d}M/2$, according to
\begin{equation}
\label{phiL}
\Phi(L) = \int_{0}^{\infty} \Phi(L \vert M) \, n(M) \, {\rm d}M.
\end{equation}
We use this equation  to  place constraints  on  $\Phi(L \vert M)$  by
using  the halo mass function  given by currently favored cosmological
models and the observed galaxy LF from the  2dFGRS.  It is easy to see
that there  is an infinite amount of  different $\Phi(L \vert M)$ that
result in exactly  the same $\Phi(L)$  for given $n(M)$.  We thus need
additional constraints, for which we use  the luminosity dependence of
the  galaxy-galaxy  two-point  correlation  function $\xi_{\rm gg}(r)$
obtained  from the 2dFGRS by  Norberg \etal (2001a).   In addition, we
use $\Phi(L \vert M) {\rm d}L$ to compute the average total luminosity
of galaxies in a halo of mass $M$,
\begin{equation}
\label{LofM}
\langle L \rangle(M) = \int_{0}^{\infty} \Phi(L  \vert M) \, L \, {\rm
d}L
\end{equation}
and compare the resulting  average mass-to-light ratios as function of
halo mass,  $\langle M/L \rangle(M)$, with  observational estimates on
the  scales  of galaxy  clusters  and  individual  galaxies (from  the
Tully-Fisher relation).  This allows  us to break the model degeneracy
and to place  interesting constraints both on galaxy  formation and on
cosmology.

This paper is organized as  follows. In Section~\ref{sec:obs} we first
give  an  overview of   the  observational  constraints used  in  this
paper. Section~\ref{sec:model} presents  our model for the conditional
luminosity  function, and  the    method used to    compute luminosity
functions, mass-to-light ratios and correlation lengths as function of
luminosity.   Results   for the   $\Lambda$CDM  concordance  model are
presented in Section~\ref{sec:res}.  In Section \ref{sec:obsuncertain}
we discuss possible  systematic  errors in current  observational data
that may  affect  our results.   Section~\ref{sec:cosmology}  presents
models  for different cosmologies and  for different assumptions about
the  nature of  the dark  matter.  A detailed  comparison with results
obtained from semi-analytical models for  the formation of galaxies is
presented  in Section~\ref{sec:sams}.    In Section~\ref{sec:light} we
use the  conditional  LF to  present several  statistics regarding the
distribution of  light in  the local  Universe, and  we  summarize our
results in Section~\ref{sec:concl}.

Throughout this  paper we define  $M$ to be  the halo mass  inside the
radius  $R_{\rm vir}$  inside of  which the  average density  is $180$
times  the   cosmic  mean  density.   All  luminosities   are  in  the
photometric  $b_j$-band  and we  convert  all  units  to $h=1$,  where
$h=H_0/(100 \kms)$.

\section{Observational Constraints}
\label{sec:obs}

\subsection{The galaxy luminosity function}
\label{sec:lf}

Galaxy  LFs are typically  obtained from  large redshift  surveys, the
size  and accuracy  of which  have improved  steadily over  the years.
(e.g., Efstathiou, Ellis \& Peterson 1988; Loveday \etal 1992; Marzke,
Huchra \& Geller 1994; Zucca \etal 1997; Folkes \etal 1999).  The most
recent determinations of  the LF of ``field'' galaxies  are those from
the Las Campanas Redshift Survey (LCRS; Lin \etal 1996), from the SDSS
commissioning data (Blanton \etal  2001), and from the 2dFGRS (Norberg
\etal 2001b).   In this paper  we use the  LF of the 2dFGRS,  which is
currently the  largest published redshift survey,  including more than
110500  galaxies  with $17.0  <  b_j <  19.2$  and  $z< 0.25$.   After
applying  $k$-corrections,   correcting  the  magnitudes   for  galaxy
evolution and  zero-point offsets, and correcting  the measured number
densities for incompleteness, Norberg \etal derive a $b_j$ LF spanning
the magnitude  range $-22.5 \lta M_{b_j}  - 5 {\rm log}  h \lta -14.0$
which is well fit by a Schechter (1976) function
\begin{equation}
\Phi(L) {\rm d}L = {\Phi^{*} \over L^{*}} \, \left({L\over
L^{*}}\right)^{\alpha} \, {\rm exp}(-L/L^{*}) \, {\rm d}L,
\end{equation}
with  $L^{*}=9.64  \times 10^{9}  h^{-2}  \Lsun$  (where we  have used
$M_{\odot,   b_j}=5.3$), $\alpha  =-1.21$   and $\Phi^{*}=1.61  \times
10^{-2}  h^{3} \Mpc^{-3}$. The  derivation of a luminosity function is
cosmology dependent, and  the parameters  listed above  correspond  to
$\Omega_0=0.3$ and  $\Omega_{\Lambda}=0.7$  \footnote{We also use  LFs
derived  for   other cosmological  models,   which   were kindly  made
available  to  us  by   Shaun   Cole   and  Peder Norberg     (private
communications).}.
 
\subsection{The galaxy correlation function}
\label{sec:corr}

Numerous   observational  studies  in   the  past   have  investigated
differential  galaxy clustering  as function  of  luminosity. Although
several of these  have claimed a dependence of  clustering strength on
luminosity (e.g., B\"orner, Mo \&  Zhou 1989; Park \etal 1994; Benoist
\etal  1996;  Willmer \etal  1998;  Guzzo  \etal  2000), others  found
results  that indicated  no significant  luminosity  dependence (e.g.,
Loveday \etal 1995; Szapudi \etal 2000; Hawkins \etal 2001).  The main
reason  for this  disagreement is  the  relatively small  size of  the
redshift surveys  used.  Recently, Norberg  \etal (2001a) investigated
the luminosity dependence of galaxy clustering using the 2dFGRS, which
is  the  largest  extant  redshift  survey.   After  correcting  their
luminosities for  band-shifting ($k$-correction) and  galaxy evolution
they computed the two-point correlation function $\xi_{\rm gg}(r)$ for
various  bins  in  absolute  magnitude.   Each  of  these  correlation
functions  is  well  fit   by  a  simple  power-law  $\xi_{\rm  gg}(r)
=(r_0/r)^{\gamma}$.   Although the shape  of the  correlation function
varies little with  luminosity (i.e., $\gamma \simeq 1.7  \pm 0.1$ for
all  bins), Norberg  \etal find  a clear  increase of  the correlation
length  $r_0$  with  increasing  luminosity.  The  correlation  length
$r_0(L)$ changes slowly  for $L \lta L^{*}$, but  increases by about a
factor two from $L=L^{*}$ to $L \simeq 5 L^{*}$.

Note that  the conversion   of   redshift into  comoving  distance  is
cosmology dependent.  The  $r_0(L)$ and $\xi_{\rm  gg}(r)$ that we use
here,  and which we  adopt  from Norberg \etal  (2001a), are  computed
assuming   $\Omega_0=0.3$  and   $\Omega_{\Lambda}=0.7$.    Note that,
although the  galaxy  luminosities have been corrected  for evolution,
the  correlation  function itself may also  evolve   with redshift, an
effect  that has not been  corrected for.  As we   show below, for the
redshift range covered by the 2dFGRS, this effect is not negligible.

\begin{figure*}
\centerline{\psfig{figure=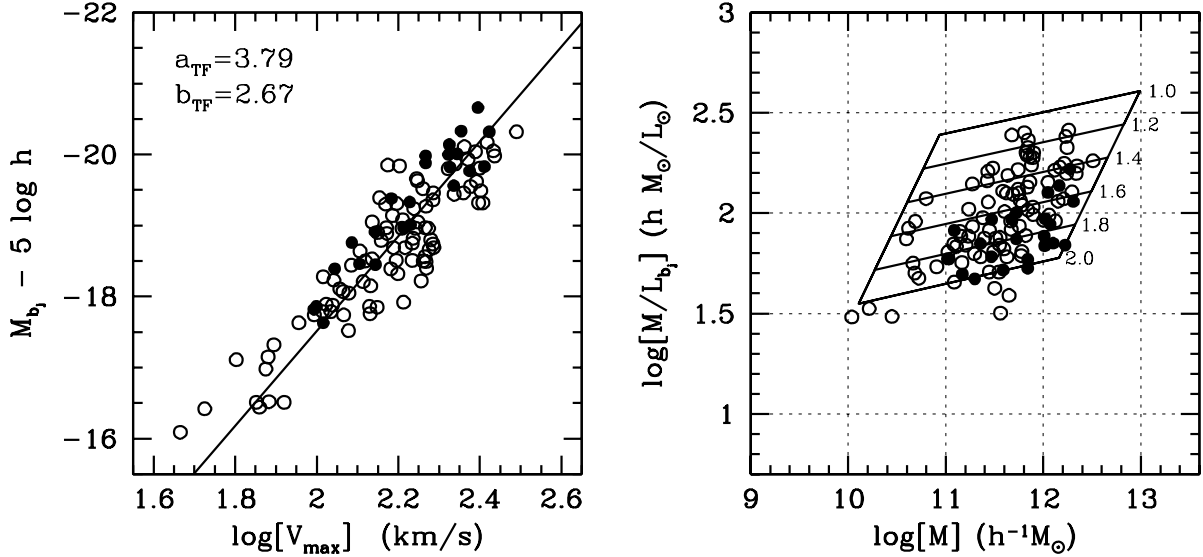,width=0.9\hdsize}}
\caption{The  left  panel   plots  the  $b_j$-band  Tully-Fisher  (TF)
relation  for the `local  calibrator sample'  (solid circles)  and the
`cluster sample'  (open circles) uncorrected  for intrinsic extinction
(data taken from  Tully \& Pierce 2000). The  solid line indicates the
best-fit linear  regression ${\rm log} L_{b_j}  = 3.79 +  2.67 \, {\rm
log}  V_{\rm  max}$.  The  panel  on  the  right plots  the  resulting
mass-to-light ratios as  function of halo mass, where  we have adopted
$V_{\rm  max}/V_{\rm   vir}=1.6$.   In  addition,   the  parallelogram
indicates the  average mass-to-light ratios of  disk galaxies obtained
from the TF  relation for six different values  of $V_{\rm max}/V_{\rm
vir}$ as indicated. See text for a more detailed discussion.}
\label{fig:tulfis}
\end{figure*}

\subsection{Tully-Fisher relation}
\label{sec:tf}

The  Tully-Fisher   (TF)  relation  for  spiral   galaxies  relates  a
luminosity measure to a  dynamical mass measure and therefore provides
an observational constraint on  the mass-to-light ratio for this class
of galaxies. The TF relation can be written as
\begin{equation}
\label{tfr}
\log(L) = a_{\rm TF} + b_{\rm TF} \, \log V_{\rm max}
\end{equation}
where  $V_{\rm  max}$  corresponds  to  the maximum  of  the  observed
rotation curve  and $a_{\rm TF}$  and $b_{\rm TF}$ are  two constants.
This can be  recast in an expression relating  the total mass-to-light
ratio to the total mass:
\begin{eqnarray}
\label{mlTF}
{\rm log} \left( {M \over L} \right) & = &
\left( 1 - {b_{\rm TF} \over 3} \right) \, \log(M) +
\left(1.797 \, b_{\rm TF} - a_{\rm TF} \right) \nonumber \\
 & & -  b_{\rm TF} \, {\rm log}  \left( {V_{\rm max}
\over V_{\rm vir}} \right) - {b_{\rm TF} \over 6} \, {\rm log} \Omega_0
\end{eqnarray}
The third term on the right hand side of equation~(\ref{mlTF}) relates
the observed rotation measure to  the virial velocity $V_{\rm vir}$ of
the dark matter halo.  Here  $V_{\rm vir}$ is the circular velocity at
the virial radius within which the average overdensity is $180$.

We  use the  data obtained  and collected  by Tully  \&  Pierce (2000;
hereafter TP00).  These authors  list photometry and HI linewidths for
24  nearby galaxies for  which distances  are available  from Cepheids
(the `local calibrator'  sample), as well as similar  data for a large
number  of  galaxies  in  various  clusters  (the  `cluster'  sample).
Distances  to these  clusters are  obtained  by assuming  that the  TF
relation  is a  universal  relation.  For  a  sample of  in total  115
galaxies  TP00  list  the  absolute  $B$-band magnitudes  and  the  HI
linewidths.   Note  that  these  magnitudes have  been  corrected  for
inclination- and linewidth-dependent  internal dust extinction.  Since
the  luminosities used  in  the LF  function  and correlation  lengths
described above have {\it not} been corrected for internal extinction,
we should also use a TF relation that has {\it not} been corrected for
the internal dust obscuration.  To  do this, we use the total apparent
$B$-band magnitudes and distance moduli listed in TP00 and compute the
absolute  $B$-band   magnitudes  corrected  for   galactic  foreground
extinction  (based  on  the   100  $\mu$m  cirrus  maps  of  Schlegel,
Finkbeiner  \&  Davis  1998).    These  are  converted  to  $b_j$-band
magnitudes using $b_j = B - 0.28 \, (B-V)$ (Blair \& Gilmore 1982) and
adopting $B-V=0.7$, which corresponds  roughly to the average color of
disk galaxies  (de Jong 1996).  Finally we introduce the  scaling with
$H_0$ adopting $h=0.7$,  which corresponds to the value  of the Hubble
constant used throughout this paper.

The  resulting $b_j$-band TF  relation is shown in   the left panel of
Figure~\ref{fig:tulfis}, where we have made the assumption that the HI
linewidth is  equal to twice   the maximum circular velocity,  $V_{\rm
max}$. Using  a  double regression with errors  in  both magnitude and
$V_{\rm max}$, we obtain the best-fit TF relation $M_{b_j} - 5 \, {\rm
log} h = -4.17 - 6.67 \, {\rm log} V_{\rm max}$, for which $a_{\rm TF}
= 3.79$  and $b_{\rm TF} = 2.67$.   These are the values we substitute
in equation~(\ref{mlTF}) to constrain  the mass-to-light ratios of our
models.

For a dark matter halo with a NFW density distribution (Navarro, Frenk
\&  White 1997)  the   circular velocity  increases  from  the center,
reaches a maximum  $V_{\rm max}$ at  $r \simeq 2.163 r_s$ (where $r_s$
is the scale radius  in the NFW  profile) and then decreases again  to
reach $V_{\rm vir}$ at the virial radius $R_{\rm vir} = c \, r_s$. One
finds that
\begin{equation}
\label{VratNFW}
{V_{\rm max}  \over V_{\rm vir}}  \simeq 0.465 \sqrt{c \,  (1+c) \over
(1+c) \, {\rm ln}(1+c) - c}
\end{equation}
With      our  definition     of      the    virial     radius    (see
Section~\ref{sec:intro}), typical galaxy sized  haloes have $10 \lta c
\lta 50$ (Jing 2000; Bullock  \etal 2001b), corresponding to $1.2 \lta
V_{\rm max}/V_{\rm vir} \lta 1.9$\footnote{Note that the $c$ used here
is not the same as that in Navarro, Frenk \& White (1997), Jing (2000)
or  Bullock \etal (2001b),    because of our  different definition  of
$R_{\rm vir}$.  When computing  the range of  expected  $c$ we use the
results of Bullock \etal (2001b) properly  scaled to our definition of
the virial radius.} The maximum rotation velocity of a  disk in a halo
is typically  larger than the $V_{\rm  max}$ of  the dark matter halo,
because  of  the contribution  of the  baryons  to  the total circular
velocity and  the  contraction  of  the dark matter  halo  during  the
formation of  the  disk (see e.g.,  Barnes  \& White 1984;  Blumenthal
\etal 1986; Flores \etal 1993; Mo, Mao \&  White 1998; Mo \& Mao 2000;
van den Bosch 2002).

In  the    right panel  of    Figure~\ref{fig:tulfis},   we  plot  the
mass-to-light ratios for the individual galaxies of  the TP00 data set
(open and  solid circles) computed   using $V_{\rm max}/V_{\rm vir}  =
1.6$.  In   addition,  we plot  the  relations between  $M/L$ and  $M$
obtained from  the best-fit linear  TF relation for 6 different values
of   $V_{\rm  max}/V_{\rm vir}$  as  indicated.    We  only plot these
relations over  the range of masses  that  correspond to the magnitude
range over which the  TF relation of TP00 has  been derived.   In what
follows we use these  TF constraints to judge  the credibility  of our
models,  keeping in mind that for  typical CDM cosmogonies one expects
$1.4 \lta  V_{\rm  max}/V_{\rm vir}  \lta  2.0$ (taking the `boosting'
effect of the baryons into account).

\section{Populating Dark Haloes with Galaxies}
\label{sec:model}

\subsection{The dark matter mass function}
\label{sec:mf}

The mass function of dark matter haloes at $z=0$ can be written in the
form
\begin{equation}
\label{halomf}
n(M) \, {\rm d}M = {\bar{\rho} \over M^2} \nu f(\nu) \,
\left| {{\rm d} {\rm ln} \sigma \over {\rm d} {\rm ln} M}\right|
{\rm d}M,
\end{equation}
where  $\bar{\rho}$ is  the mean  matter  density of  the Universe  at
$z=0$,  $\nu   =  \delta_c/\sigma(M)$,  $\delta_c$   is  the  critical
overdensity required for collapse at $z=0$, and $f(\nu)$ is a function
of $\nu$ to be specified below.  The quantity $\sigma(M)$ in the above
equation is the  linear rms mass fluctuation on  mass scale $M$, which
is given by the linear  power spectrum of density perturbations $P(k)$
as
\begin{equation}
\label{variance}
\sigma^2(M) = {1 \over 2 \pi^2} \int_{0}^{\infty} P(k) \;
\widehat{W}_{M}^2(k) \; k^2 \; {\rm d}k,
\end{equation}
where $\widehat{W}_{M}(k)$  is the Fourier transform  of the smoothing
filter on mass scale $M$.

According to the Press-Schechter formalism  (Press \& Schechter 1974),
the function $f(\nu)$ has the universal form,
\begin{equation}
\label{fnuPS}
\nu f(\nu) = 2\left({\nu^2\over 2\pi}\right)^{1/2}
\exp\left(-{\nu^2\over 2}\right),
\end{equation}
independent   of  cosmology,  redshift,   power   spectrum   and   its
normalization (e.g.,  Bond \etal 1991;  Lacey \& Cole 1993).  However,
various  studies have  shown  that the  halo mass  function  with this
$f(\nu)$ is inconsistent with numerical simulations (e.g., Efstathiou,
Ellis \& Peterson 1988; Jain \&  Bertschinger 1994; Tormen 1998; Gross
\etal  1998;   Governato  \etal  1999;  Jenkins  \etal   2001).  Using
ellipsoidal  rather than spherical collapse   conditions, Sheth, Mo \&
Tormen (2001) derived an improved form for $f(\nu)$ given by
\begin{equation}
\label{fnuST}
\nu \, f(\nu) = 2 A \,\left(1 + {1\over \nu'{^{2q}}}\right)\
\left({\nu'{^2}\over 2\pi}\right)^{1/2}
\exp\left(-{\nu'{^2} \over 2}\right)\,
\end{equation}
with  $\nu'=\sqrt{a}\,\nu$,  $a=0.707$, $q=0.3$ and  $A\approx 0.322$.
The   resulting  mass function  has  been  shown   to be in  excellent
agreement    with numerical simulations, as   long  as halo masses are
defined as the  masses inside a sphere with  an average overdensity of
about $180$ (Sheth \& Tormen 1999; Jenkins \etal 2001).  Therefore, in
what follows we consistently use that  definition of halo mass, and we
adopt a  mass function with  $f(\nu)$ given by equation~(\ref{fnuST}).
In addition we use the CDM power spectrum of Bardeen \etal (1986), and
we adopt a spatial top-hat filter for which
\begin{equation}
\label{THfour}
\widehat{W}_{M}(k;R) =  {3 \over (k  R)^3} \left[ \sin(k R)  - k
R \cos(k R)\right]
\end{equation}
where the mass $M$ and filter radius $R$ are related according to $M
= 4 \pi \bar{\rho} R^3 / 3$.

\subsection{The conditional luminosity function}
\label{sec:condLF}

In order  to compute  a LF  from the halo  mass  function, we  need to
specify the conditional luminosity function $\Phi(L \vert M) {\rm d}L$
(see equation~[\ref{phiL}]),    which  gives the   expected  number of
galaxies with  luminosities in the range  $L \pm {\rm d}L/2$  (in some
chosen  photometric band) in  a halo of mass   $M$.  Note that $\Phi(L
\vert M)$ is a statistical function,  and should not be interpreted as
the LF of galaxies residing in any {\it individual} dark matter halo.
\begin{figure}
\centerline{\psfig{figure=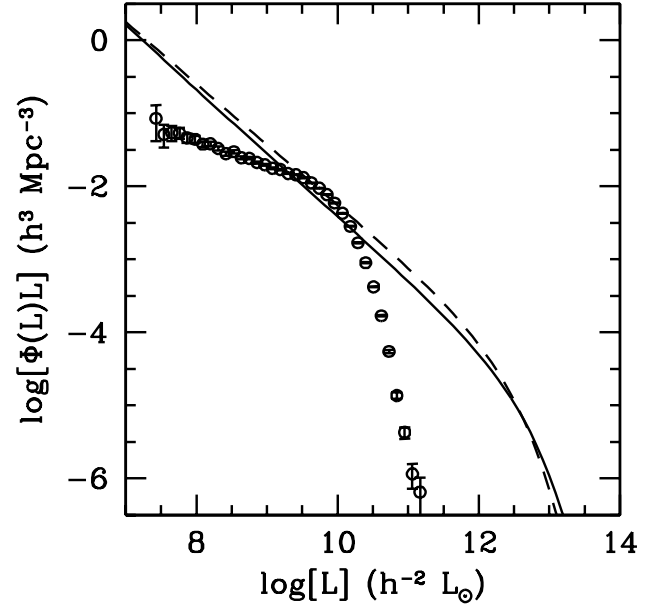,width=\hssize}}
\caption{A  comparison of the galaxy  LF with  the halo mass function.
Open circles with errorbars   correspond  to the   2dFGRS LF  in   the
$b_j$-band. Solid (dashed)  lines correspond to  the LF that one would
obtain from  the Sheth \& Tormen  (Press-Schechter) halo mass function
under the assumption  that each halo yields   exactly one galaxy  with
$M/L = 100  h  \MLsun$. Note that   under such naive assumptions   one
expects  too many both faint  and  bright galaxies, suggesting that in
reality  the $M/L$ decreases (increases) with  mass  at the low (high)
mass end.}
\label{fig:lf}
\end{figure}

For massive  haloes, such as  clusters of galaxies, that  contain many
galaxies the shape  of $\Phi(L\vert M)$ should be the  same as that of
the cluster LF,  which can be well described  by a Schechter function.
For all  mass haloes,  the average of  $\Phi(L\vert M)$ over  the halo
mass function should  give the field galaxy LF.   Therefore, we assume
that  $\Phi(L\vert  M) {\rm  d}L$  can  be  described by  a  Schechter
function:
\begin{equation}
\label{phiLM}
\Phi(L \vert M) {\rm d}L = {\tilde{\Phi}^{*} \over \wLstar} \,
\left({L \over \wLstar}\right)^{\walpha} \,
\, {\rm exp}(-L/\wLstar) \, {\rm d}L
\end{equation}
Here   $\wLstar   =    \wLstar(M)$,   $\walpha   =   \walpha(M)$   and
$\tilde{\Phi}^{*} =  \tilde{\Phi}^{*}(M)$; i.e., the  three parameters
that describe the conditional LF depend on $M$.  In what follows we do
not explicitly write this  mass dependence, but consider it understood
that quantities with a tilde are functions of $M$.

With $\Phi(L  \vert M)$  defined by equation~(\ref{phiLM}),  the total
average luminosity in a halo of mass $M$ is
\begin{equation}
\label{meanL}
\langle L \rangle(M) = \int_{0}^{\infty}  \Phi(L \vert M) \, L \, {\rm
d}L = \tilde{\Phi}^{*} \, \wLstar \, \Gamma(\walpha+2)
\end{equation}
with $\Gamma(x)$  the Gamma function.  The average  number of galaxies
brighter than $\wLstar$ in a halo of mass $M$ is
\begin{equation}
\label{meanN}
{\cal N}^{*}(M) \equiv \int_{\wLstar}^{\infty} \Phi(L \vert M) \, {\rm
d}L = \tilde{\Phi}^{*} \, \Gamma(\walpha+1,1).
\end{equation}
with $\Gamma(a,x)$ the incomplete Gamma function.

For each halo, we define a `central' galaxy whose luminosity we denote
by $L_c$.  We assume the  central galaxy to be the  brightest one in a
halo, consistent with the  fact that in most  (if not all) haloes  the
brightest  members  reside near   the center. The  luminosity of  this
central galaxy is defined as
\begin{equation}
\label{Lcentral}
L_c(M) = \int_{L_1}^{\infty}  \Phi(L \vert M) \, L \, {\rm
d}L = \tilde{\Phi}^{*} \, \wLstar \, \Gamma(\walpha+2,L_1/\wLstar),
\end{equation}
with  $L_1$ defined so  that a  halo  of mass $M$   has on average one
galaxy with $L > L_1$, i.e.,
\begin{equation}
\label{Lone}
\int_{L_1}^\infty \Phi(L\vert M) dL=1\,.
\end{equation}
Note  that  with this  definition,  the  luminosities  of the  central
galaxies in  haloes of the same mass  are all the same.   We have also
experimented  with  drawing  $L_c$  at random  from  the  distribution
function  $\Phi(L\vert M)$  at $L>L_1$,  but found  that this  did not
significantly influence our results. We therefore adopt the definition
of equation~(\ref{Lcentral}) throughout.  This is further motivated by
numerical simulations and  semi-analytical models of galaxy formation,
which suggest  that the luminosities  of central galaxies  are tightly
correlated with the masses of  their host haloes (e.g.  Katz, Weinberg
\& Hernquist  1996; Fardal  \etal 2001; Pearce  \etal 2000;  Kay \etal
2001;  Kauffmann \etal 1993;  Somerville \&  Primack 1999;  Cole \etal
2000).
 
In order to  fully specify the conditional LF, we  need to specify the
mass dependence of $\tilde{\Phi}^{*}$, $\wLstar$ and $\walpha$. We are
guided by a direct comparison of the halo mass function $n(M)$ and the
galaxy LF, $\Phi(L)$. Under the  assumption that each dark matter halo
harbors exactly one galaxy, and  that each system has exactly the same
mass-to-light  ratio $M/L$, the  galaxy LF  follows directly  from the
halo mass function.  Figure~\ref{fig:lf} compares the LF thus obtained
for $M/L =  100 h \MLsun$ with  that of the 2dFGRS.  The  actual LF is
steeper (shallower) than the one  obtained directly from the halo mass
function at  high (low) luminosities.  This  immediately suggests that
in  reality,  rather  than   being  constant,  $\langle  M/L  \rangle$
decreases with  increasing mass, reaches a minimum  at around $L^{*}$,
and then increases again.   We therefore consider the following simple
parameterization,
\begin{equation}
\label{MtoLmodel}
\left\langle {M \over L} \right\rangle (M) = {1 \over 2} \,
\left({M \over L}\right)_0
\left[ \left({M \over M_1}\right)^{-\beta} +
\left({M \over M_1}\right)^{\gamma_1}\right],
\end{equation}
which has four free parameters: a characteristic mass $M_1$, for which
the mass-to-light ratio is equal to $(M/L)_0$, and two slopes, $\beta$
and $\gamma_1$, which specify the behavior of $\langle M/L \rangle$ at
the low and high mass ends, respectively. For $M \ll M_1$ one has that
$\langle M/L  \rangle \propto M^{-\beta}$,  while for $M \gg  M_1$ the
mass-to-light  ratio increases  with  mass according  to $\langle  M/L
\rangle \propto M^{\gamma_1}$ (both $\beta$ and $\gamma_1$ are assumed
to, but not restricted to, be positive).

In addition we introduce a  fifth free parameter, $M_{\rm min}$, which
sets  the  mass scale  below  which no  galaxies  form;  i.e., we  set
$\langle L \rangle(M) = 0$ for $M < M_{\rm min}$. This additional free
parameter has  a physical motivation  (see below), and is  included to
take account  of the  fact that  the 2dFGRS LF  has a  faint magnitude
cut-off at $M_{b_j} - 5 {\rm log} h = -14.0$.

It is  obvious that  for galaxy sized  haloes $\wLstar$  must decrease
with  decreasing mass,  since galaxies  with $L=10^{10}  h^{-2} \Lsun$
should not be found in  a halo of $M=10^9 h^{-1} \Msun$.  Furthermore,
since the LF of the total galaxy population has an exponential tail at
$L>L^{*}$, it  is unlikely that  $\wLstar$ keeps increasing  with $M$,
because otherwise  we would observe galaxies brighter  than, say, $100
L^{*}$.   Therefore  we   adopt  the  following  parameterization  for
$\wLstar(M)$:
\begin{equation}
\label{LstarM}
{M \over \wLstar(M)} = {1 \over 2} \, \left({M \over L}\right)_0 \,
f(\walpha) \, \left[ \left({M \over M_1}\right)^{-\beta} +
\left({M \over M_2}\right)^{\gamma_2}\right],
\end{equation}
with
\begin{equation}
\label{falpha}
f(\walpha) = {\Gamma(\walpha+2) \over \Gamma(\walpha+1,1)}.
\end{equation}
This  parameterization   of  $\wLstar(M)$  has   two  additional  free
parameters:  a  characteristic  mass   $M_2$  and  a  power-law  slope
$\gamma_2$.   The  factor  $f(\walpha)$   is  to  ensure  that  ${\cal
N}^{*}(M)=1$ for $M \ll {\rm min}[M_1,M_2]$, which is motivated by the
fact that for  small mass haloes (i.e., smaller  than a typical galaxy
group) one expects  on average only one dominant  galaxy per halo. For
$M \gg {\rm max}[M_1,M_2]$ this parameterization yields ${\cal{N}}^{*}
\propto M^{\gamma_2 - \gamma_1}$.

Finally,  for  $\walpha(M)$ we  adopt  a  simple  linear scaling  with
$\log(M)$, which we parameterize as follows:
\begin{equation}
\label{alphaM}
\walpha(M) = \alpha_{15} + \eta \, \log(M_{15}).
\end{equation}
Here  $M_{15}$  is  the  halo  mass  in  units  of  $10^{15}  \msunh$,
$\alpha_{15} = \walpha(M_{15}=1)$, and  $\eta$ describes the change of
the  faint-end  slope  $\walpha$  with  halo  mass.   Note  that  once
$\walpha(M)$  and $\wLstar (M)$  are given,  the normalization  of the
conditional LF,  $\tilde{\Phi}^{*} (M)$  can be specified  through the
mass-to-light       ratio       using      equations~(\ref{MtoLmodel})
and~(\ref{meanL}).

Our  model  for  $\Phi(L  \vert  M)$  specified  above 
thus has  $9$  free
parameters:   $(M/L)_0$,  $M_{\rm   min}$,   $M_1$,  $M_2$,   $\beta$,
$\gamma_1$,  $\gamma_2$,  $\eta$ and  $\alpha_{15}$.   Clearly, it  is
infeasible  to  explore the  entire  $9$-dimensional parameter  space.
However,  we   can  use  observational   constraints  and  theoretical
considerations to significantly reduce the freedom of the model. First
of  all,  we use  constraints  obtained from  studies  of  the LFs  of
clusters of galaxies.  Beijersbergen \etal (2002) derived the $B$-band
LF of the Coma cluster, which is well fit by a Schechter function with
$L_{\rm coma}^{*}  = 6.73 \times  10^9 h^{-2} \Lsun$  and $\alpha_{\rm
coma} = -1.32$.  Therefore,  we set $\alpha_{15}=-1.32$, which reduces
the number of free parameters  by one. Secondly, $L_{\rm coma}^{*}$ is
very similar  to $L^{*}$  of the {\it  entire} galaxy  population (see
Section~\ref{sec:obs}).   In  addition,  the luminosity  functions  of
galaxies  in  groups  and   poor  clusters  also  have  characteristic
luminosities similar to $L^{*}$ (e.g. Muriel, Valotto \& Lambas 1998).
This suggests that $\wLstar$ does not increase much with mass over the
range $10^{13} -  10^{15} \msunh$, and we therefore  tune the value of
$M_2$ so that  $\wLstar(M_L)=L^{*}$ with $M_L=10^{13.5} \msunh$. Based
on a variety  of observations, Fukugita, Hogan \&  Peebles (1998) find
that  clusters  of  galaxies,   which  they  define  as  systems  with
$M>10^{14} h^{-1}  \Msun$, have a  mass-to-light ratio $(M/L)_B  = 450
\pm 100 h \MLsun$.  Therefore, we  set $\langle M/L\rangle (M) = 500 h
\MLsun$ for $M \geq 10^{14} h^{-1} \Msun$, and tune $\gamma_1$ so that
$\langle  M/L \rangle  (M)$  reaches this  upper  limit at  $M=10^{14}
h^{-1} \Msun$. Finally we adopt  $M_{\rm min} = 1.0 \times 10^9 h^{-1}
\Msun$, which  corresponds roughly  to the mass  scale below  which no
galaxies can form in a re-ionized medium, i.e., the virial temperature
is $T_{\rm vir} \lta 10^{4}$K for such systems at the present time.

All in all we have thus reduced the number of free parameters to five:
$(M/L)_0$,    $M_1$,   $\beta$,    $\gamma_2$,    and   $\eta$.     In
Section~\ref{sec:moduncertain}  we   address  the  impact   that  this
reduction of the model freedom has on our results.

\subsection{The galaxy correlation function}
\label{sec:galcorr}

If  the initial  density distribution  of   matter is  described by  a
Gaussian random field, the clustering properties of dark matter haloes
are fully determined by the linear power spectrum $P(k)$ and cosmology
(Kaiser  1984; Bardeen \etal  1986; Cole \&  Kaiser  1989; Mo \& White
1996; Mo, Jing \& White 1997; Sheth, Mo \& Tormen 2001).  Once a model
is  adopted for the galaxy  occupation  numbers of  dark matter haloes
(i.e., once $P(N \vert M)$ is specified), the clustering properties of
galaxies, such    as  the two-point  correlation    function $\xi_{\rm
gg}(r)$,  can   be  computed.  With our   model   for  the conditional
luminosity function  $\Phi(L \vert  M)$ described above,  we basically
have completely specified the  differential halo occupation numbers as
function of luminosity and  halo mass.   Therefore,  once we adopt  an
additional  model    for the  spatial   distribution  of  galaxies  in
individual haloes, we can compute the correlation function of galaxies
{\it as function of  luminosity}.  Using the definition that $\xi_{\rm
gg}(r_0)=1$ we compute $r_0(L)$,  which  we  compare to the   observed
relation of the 2dFGRS (see Section~\ref{sec:corr}).

We  proceed  as follows.   We  decompose  the two-point  galaxy-galaxy
correlation function into two parts:
\begin{equation}
\label{ggcorrsplit}
\xi_{\rm gg}(r) = \xi_{\rm gg}^{1 {\rm h}}(r) +
                  \xi_{\rm gg}^{2 {\rm h}}(r) \, .
\end{equation}
Here $\xi_{\rm  gg}^{1 {\rm  h}}$ represents  the  correlation due  to
pairs of galaxies  within the same halo  (the ``1-halo''  term), while
$\xi_{\rm  gg}^{2 {\rm h}}$  describes the correlation due to galaxies
that occupy  different  haloes (the ``2-halo''  term). For  the 1-halo
term we  need to specify  the  distribution of galaxies in  individual
haloes,  while the  2-halo  term is given by   the correlation of  the
population  of  dark matter  haloes.    

\subsubsection{The 2-halo term}
\label{sec:twohalo}

Consider all galaxies with luminosities in the range $[L_1, L_2]$, and
define  $\hxi_{\rm gg}(r)$  as the  two-point correlation  function of
this subset of  galaxies. By definition, the total  number of pairs of
galaxies {\it per comoving volume}  in this luminosity range that have
separations in the range $r  \pm {\rm d}r/2$ is
\begin{equation}
\label{corrdef}
n_{\rm  pair} =  {\overline{n}_{\rm g}^2  \over 2}  \, [1  + \hxi_{\rm
gg}(r)] \, 4 \pi r^2 {\rm d}r.
\end{equation}
where the factor $1/2$ corrects for double counting of each pair. Here
$\overline{n}_{\rm g}$ is the mean  number density of galaxies with $L
\in [L_1, L_2]$ which is given by
\begin{equation}
\label{barng}
{\overline n}_{\rm g} = \int_{0}^{\infty} n(M) \, \langle N(M)\rangle
\, {\rm d}M \,,
\end{equation}
and
\begin{equation}
\label{nlm}
\langle N(M) \rangle = \int_{L_1}^{L_2} \Phi (L\vert M) \, {\rm d}L
\end{equation}
gives the  mean number of  galaxies in the specified  luminosity range
for haloes of mass $M$.
\begin{figure}
\centerline{\psfig{figure=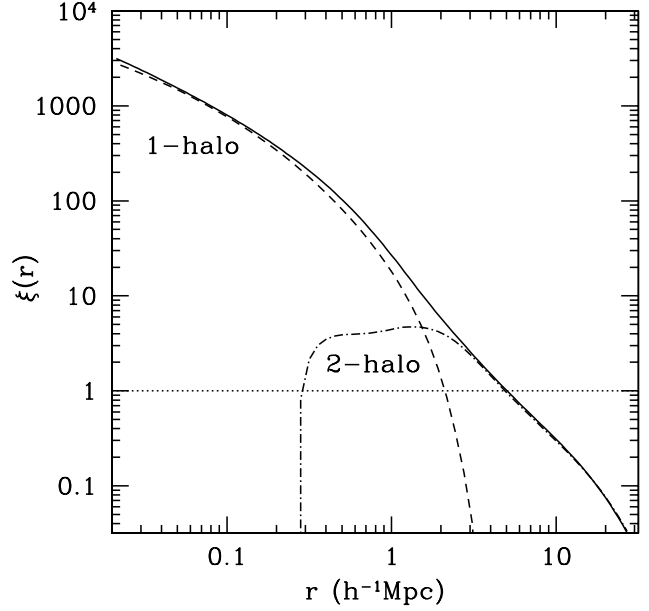,width=\hssize}}
\caption{The   non-linear  dark   matter  mass   correlation  function
$\xi_{\rm  dm}(r)$ of  the $\Lambda$CDM  ``concordance''  model (solid
line).  The  dashed and dot-dashed  lines indicate the  ``1-halo'' and
``2-halo''  terms,  respectively.  Note  that  the correlation  length
$r_0$, defined  as $\xi_{\rm dm}(r_0)=1$ (dotted  line), is determined
by the ``2-halo'' term only.}
\label{fig:xi}
\end{figure}

Consider a galaxy with $L \in [L_1, L_2]$ that lives in a halo of mass
$M_1$.  There  are, on average, $n(M_2)  \, 4 \pi r^2  {\rm d}r$ dark
matter haloes with mass $M_2$ that are separated from this galaxy by a
distance in  the range $r  \pm {\rm d}r/2$.   In each of  these haloes
there are on average $\langle  N(M_2) \rangle$ galaxies that also have
a luminosity in the range $[L_1,L_2]$.  Integrating over the halo mass
function,  and taking  account of  the clustering  properties  of dark
matter haloes,  we obtain  the {\it excess}  number $N_{+}$  of galaxy
pairs {\it per galaxy in a halo of mass} $M_1$ with separations in the
range $r\pm {\rm d}r/2$:
\begin{eqnarray}
\label{nexc}
\lefteqn{N_{+}(M_1) = 4 \pi r^2 {\rm d}r} \nonumber \\
& & \times \int_{0}^{\infty} n(M_2) \,\langle N(M_2)\rangle \, \xi_{\rm hh}(r;
M_1, M_2) \, {\rm d}M_2.
\end{eqnarray}
Here $\xi_{\rm hh}(r; M_1, M_2)$ is the cross correlation between dark
matter haloes of mass $M_1$  and $M_2$. Note that we implicitly assume
that all  galaxies in  a given  halo are located  at the  halo center.
Since  the  separation between  individual  haloes  is typically  much
larger than  the separation  between galaxies in  the same  halo, this
simplification does not influence our results for large $r$.

Multiplying $N_{+}(M_1)$ with $\langle N(M_1) \rangle$ and integrating
over the  halo mass function gives  the total excess  number of galaxy
pairs  {\it per  comoving  volume} with  separations  in the  required
range.  Combining this with equation~(\ref{corrdef}) yields
\begin{equation}
\label{fbarn}
{{\overline n}_{\rm g}^2 \over 2} \, \hxi_{\rm gg}^{\rm 2h}(r) \, 
4 \pi r^2dr = {1 \over 2} \int_{0}^{\infty} n(M_1) 
\langle N(M_1) \rangle \, N_{+}(M_1) \, {\rm d}M_1.
\end{equation}

What  remains  is to specify     the dark halo  correlation   function
$\xi_{\rm hh}(r; M_1,M_2)$.   Mo  \& White (1996)  developed  a model,
based  on  the Press-Schechter  formalism, that  describes the bias of
dark matter  haloes of mass  $M$ with respect to  the dark matter mass
distribution. This model was extended by Sheth, Mo \& Tormen (2001) to
account for ellipsoidal collapse. According to this,
\begin{equation}
\label{xihalo}
\xi_{\rm hh}(r;M_1,M_2) = b(M_1) \, b(M_2) \, \xi_{\rm dm}^{\rm 2h}(r),
\end{equation}
where $\xi_{\rm dm}^{\rm 2h}(r)$ is the 2-halo term of the dark matter
mass correlation function (see below) and
\begin{eqnarray}
\label{bm}
b(M) & = & 1 + {1\over\sqrt{a}\delta_{\rm sc}(z)} \
\Bigl[ \sqrt{a}\,(a\nu^2) + \sqrt{a}\,b\,(a\nu^2)^{1-c} - \nonumber \\
& & {(a\nu^2)^c\over (a\nu^2)^c + b\,(1-c)(1-c/2)}\Bigr],
\end{eqnarray}
with  $a=0.707$, $b=0.5$,  $c=0.6$ and  $\nu =  \delta_c  / \sigma(M)$
(Sheth, Mo \&  Tormen 2001).  This model has  been shown to accurately
match  the correlation  function  of dark  matter  haloes in  $N$-body
simulations (Jing 1998; Sheth \& Tormen 1999).

Substituting         equations~(\ref{xihalo})         and~(\ref{nexc})
in~(\ref{fbarn}) we obtain
\begin{equation}
\label{xi2h}
\hxi_{\rm gg}^{\rm 2h}(r) = \overline{b}^2 \, \xi_{\rm dm}^{\rm 2 h}(r) ,
\end{equation}
where
\begin{equation}
\label{averbias}
\overline{b} = {1 \over \overline{n}_{\rm g}} \int_{0}^{\infty} n(M) \,
\langle N(M) \rangle \, b(M) \, {\rm d}M\,.
\end{equation}
Note  that the  dark matter  mass  correlation function  in the  above
expression is the 2-halo term (i.e., the 1-halo term is irrelevant for
the   halo-halo  correlation).    Using   the  same   concept  as   in
equation~(\ref{ggcorrsplit})  we  can  write that  $\xi_{\rm  dm}^{\rm
2h}(r) = \xi_{\rm dm}(r)  - \xi_{\rm dm}^{\rm 1h}(r)$.  Here $\xi_{\rm
dm}(r)$ is  the non-linear two-point correlation function  of the dark
matter distribution,  and $\xi_{\rm dm}^{1 {\rm h}}(r)$  is the 1-halo
term. The Fourier transform of the one-halo term is given by
\begin{equation}
\label{powspeconehalo}
P^{\rm {1h}}_{\rm dm}(k)= \int_{0}^{\infty} n(M) \,
\left[ \hat{\delta}(M;k) \right]^2 {\rm d}M,
\end{equation}
where 
\begin{equation}
\label{FTdensprof}
\hat{\delta}(M;k) =  \int_{0}^{\infty} {\rho(r) \over \bar{\rho}} \,
{\rm e}^{-i{\bf  k}\cdot{\bf r}}  {\rm d}^3{\bf r}
\end{equation}
is  the  Fourier  transform  of  the halo  density  profile  $\rho(r)$
(e.g. Neyman \& Scott 1952; McClelland  \& Silk 1977; Ma \& Fry 2000).
We assume that $\rho(r)$ has the NFW form
\begin{equation}
\label{NFW}
\rho(r) = \frac{\bar{\delta}\bar{\rho}}{(r/r_{\rm s})(1+r/r_{\rm
 s})^{2}},
\end{equation}
where $r_s$  is a characteristic  radius, $\bar{\rho}$ is  the average
density  of  the  Universe,  and  $\bar{\delta}$  is  a  dimensionless
amplitude which  can be expressed  in terms of the  halo concentration
parameter $c=R_{\rm vir}/r_s$ as
\begin{equation}
\label{overdensity}
\bar{\delta} = {180 \over 3} \, {c^{3} \over {\rm ln}(1+c) - c/(1+c)}.
\end{equation}
Numerical simulations show that $c$ is correlated  with halo mass, and
we use the relation  given by Bullock \etal  (2001b), converted to the
$c$ appropriate for our definition of $R_{\rm vir}$.

The 1-halo and  2-halo terms of the dark  matter two-point correlation
function  thus  obtained  are   shown  in  Figure~\ref{fig:xi}  for  a
$\Lambda$CDM model  with $\Omega_0=0.3$, $\Omega_\Lambda=0.7$, $h=0.7$
and  $\sigma_8=0.9$.   Here  we  have  used the  fitting  formula  for
$\xi_{\rm  dm}(r)$ given  by Peacock  \& Dodds  (1996). Note  that for
$r\ga  3\mpch$, $\xi_{\rm  dm}(r)$ is  dominated by  the  2-halo term,
while for $r\la 1\mpch$ the  1-halo term dominates.  Therefore as long
as ${\overline b}$  is sufficiently large, $r_0$ is  determined by the
2-halo term only. In this case, the value of $r_0$ depends only on the
mean occupation  number $\langle N(M)\rangle  $ and is  independent of
the details of  how galaxies are distributed in  individual haloes. In
practice, we  always find that  at $r=r_0$ the 2-halo  term dominates,
and   we   therefore   use   equation~(\ref{xi2h})  to   compute   the
galaxy-galaxy correlation function for galaxies with $L \in [L_1,L_2]$
from which we obtain  $r_0(L)$ by solving $\hxi^{\rm 2h}_{\rm gg}(r_0)
= 1$.

As  pointed out  in Section~\ref{sec:obs},  the  correlation functions
obtained by Norberg \etal (2001a) have not been corrected for possible
redshift evolution. This implies that the $r_0(L)$ measurements do not
correspond to  $z=0$.  In  fact, since more  luminous galaxies  can be
detected out  to higher redshift, the correlation  lengths of brighter
galaxies  correspond  to  galaxy  populations  with  a  higher  median
redshift.   Note that the  luminosities {\it  have} been  corrected to
$z=0$,  and the  LF  used  therefore corresponds  to  galaxies at  the
present time.  Thus,  in order to compare model  and observations in a
consistent way, we calculate the model LF at $z=0$ and the correlation
function   at  some   characteristic   redshift  of   the  sample   in
consideration.   As an  approximation, we  estimate the  mean redshift
${\overline  z}$  for  each  volume-limited sample  in  Norberg  \etal
(2001a),   and   calculate    an   effective   bias   factor   $b_{\rm
eff}({\overline z})$  at this  redshift from the  bias factor  $b$ for
$z=0$  given by  equation~(\ref{xi2h}).  Assuming  that the  number of
galaxies is conserved  in the redshift interval covered  by the 2dFGRS
($\Delta z\sim 0.3$ for the  brightest sample, and smaller for fainter
samples), one can show that these two bias factors are related by
\begin{equation}
\label{effbias}
b_{\rm eff}({\overline z}) = 1 + [D(0)/D({\overline
z})](b-1)\,, 
\end{equation}
(Mo \& White  1996, 2002; Fry 1996) where $D(z)$  is the linear growth
rate at  redshift $z$, related to  the linear growth  factor $g(z)$ by
$D(z)=g(z)/(1+z)$.   We use this  effective bias  factor and  the mass
correlation function at the mean redshift ${\overline z}$ to calculate
the correlation length $r_0$ for  each luminosity sample.  For all the
magnitude samples used  here, the effect is modest,  amounting to $\la
10\%$ change in $r_0$.
 
\subsubsection{The 1-halo term}
\label{sec:onehalo}

For small  separations, where  the number of  pairs are mostly  due to
galaxies in  individual haloes, $\xi_{\rm gg}(r)$ depends  not only on
the occupation number of galaxies, but also on how galaxies are 
distributed in individual haloes.

To model the galaxy distribution in  individual haloes, we assume that
each  halo contains  one  central galaxy (with  luminosity  $L_c$, see
equation~[\ref{Lcentral}]) which is located  at the halo center.   The
other galaxies, which we call satellite galaxies, are distributed with
a  number density profile that is  identical to the density profile of
the  dark matter (equation~[\ref{NFW}]).    This is  clearly  a simple
assumption, because processes such as  dynamical friction may lead  to
some segregation between galaxies and dark matter particles.
   
If the luminosity of the central galaxy $L_c\in [L_1, L_2]$, the total
number of pairs between the central galaxy and satellite galaxies in a
halo of mass $M$ is equal to the total number of satellite galaxies in
the  given luminosity  range,  $N_{\rm cs}=N_{\rm  s} (M)$;  otherwise
(i.e. when  $L_c$ is  not in  $[L_1, L_2]$) this  number is  zero. The
total  number  of distinct  pairs  of  satellite  galaxies is  $N_{\rm
ss}={1\over  2}N_{\rm  s}(N_{\rm  s}-1)$.   The distribution  of  pair
separations between central and satellite galaxies, which we denote by
$f_{\rm  c}(r){\rm d}r$,  has the  NFW form.   The satellite-satellite
pairs  follow  a different  distribution  of  separations, denoted  by
$f_{\rm s}(r)\,{\rm d}r$, which can be determined from the NFW density
profile.  The  total number  of galaxy pairs  with separations  in the
range $r\pm {\rm d}r/2$ can thus be written as
\begin{equation}
\label{npairM}
N_{\rm pair}(M) f(r)
= N_{\rm ss}(M)f_{\rm s}(r)+ N_{\rm cs}(M)f_{\rm c}(r)\,,
\end{equation}
where
\begin{equation}
\label{secnpairM}
N_{\rm pair}(M) = {1 \over 2} \langle N(M) (N(M)- 1) \rangle
\end{equation}
is  the total number  of  galaxy pairs  regardless of  separation, and
$f(r)$ is determined through equation~(\ref{npairM}).  The 1-halo term
of the galaxy correlation function can then be estimated from
\begin{equation}
\label{xi1h}
 \frac{{\overline n}_{\rm g}^2}{2}\xi_{\rm gg}^{\rm 1h}
(r)4\pi r^2 {\rm d}r =
{\rm d} r \int n(M) \langle N_{\rm pair}(M)\rangle f(r)\, {\rm d} M,
\end{equation}
where $\langle N_{\rm pair}(M)\rangle$ is  the mean number of pairs in
haloes  of  mass $M$  (see  Berlind \&  Weinberg  2002  for a  similar
approach).

Note that  $\langle N_{\rm  pair}(M)\rangle$ depends  not only  on the
mean occupation $\langle N(M)\rangle$,  but also on the second moment.
Thus, in general, the 1-halo term depends also on the deviation of the
occupation number from the mean. This deviation depends on the details
of  galaxy formation (e.g. Kauffmann  \etal  1999; Benson \etal 2000).
As a simple   model,  we make the    assumption that $N(M)$ has    the
probability of $N+1-\langle N(M)\rangle$ to take the value $N$ and the
probability   of $\langle N(M)\rangle-N$ to take   the value $N+1$, if
$N<\langle N(M)\rangle <N+1$.  In this case,  the mean number of pairs
is\footnote{Note:  Berlind \& Weinberg  (2002),  using the same model,
adopted $\langle N_{\rm pair} \rangle  = \langle N \rangle (\langle  N
\rangle - 1)$.   This  is only  approximately correct and  can lead to
small errors  for individual haloes with  $\langle  N \rangle \lta 5$.
For the computation  of $\xi_{\rm gg}(r)$, however, this approximation
is sufficiently accurate.}
\begin{equation}
\label{npair}
\langle N_{\rm pair} \rangle = N \, \langle N \rangle - 
{1 \over 2} \, N \, (N+1)
\end{equation}
This particular model for the distribution  of halo occupation numbers
is supported by the semi-analytical models of Benson \etal (2000), who
found that the halo  occupation  probability distribution is  narrower
than a Poisson distribution with  the same mean.  In addition, Berlind
\& Weinberg (2002) have shown that distributions narrower than Poisson
are more successful in yielding power-law correlation functions.

We use   the 1-halo   term only  to  investigate the   {\it shape}  of
$\xi_{\rm gg}(r)$, not to compute $r_0$. Since the 1-halo term depends
on, and is rather sensitive to, the assumptions about the distribution
of galaxies inside dark matter haloes, we  primarily focus on $r_0(L)$
and   only briefly discuss  the     shape  of  $\xi_{\rm gg}(r)$    in
Section~\ref{sec:corrshape}.

\section{Results}
\label{sec:res}

Having specified  the model and  the observational constraints  we now
proceed as  follows. After choosing  a cosmological model,  which sets
the  mass  function $n(M)  {\rm  d}M$  and  the two-point  correlation
function  $\xi_{\rm hh}(r)$  of  dark matter  haloes,  we specify  the
conditional LF  $\Phi(L \vert M)$  and compute the  LF $\hat{\Phi}(L)$
and   correlation   lengths  $\hat{r}_0(L)$   of   the  model,   using
equations~(\ref{phiL})  and~(\ref{xi2h}), respectively.  Note  that in
all  cases discussed  here  the correlation  lengths, defined  through
$\xi_{\rm gg}(r_0)=1$,  are completely  determined by the  2-halo term
$\xi_{\rm gg}^{\rm 2h}(r)$ [i.e., $\hat{r}_0(L)$ is independent of our
model  for the  distribution  of galaxies  in  individual dark  matter
haloes].  Using Powell's multi-dimensional direction set method (e.g.,
Press \etal  1992), we find  the values of $(M/L)_0$,  $M_1$, $\beta$,
$\gamma_2$ and $\eta$ that minimize
\begin{equation}
\label{chisq}
\chi^2 = \chi^2(\Phi)+\chi^2(r_0).
\end{equation}
Here
\begin{equation}
\label{chisqLF}
\chi^2(\Phi) = \sum_{i=1}^{N_{\Phi}}
\left[ {\hat{\Phi}(L_i) - \Phi(L_i) \over \Delta \Phi(L_i)} \right]^2,
\end{equation}
and
\begin{equation}
\label{chisqr0}
\chi^2(r_0) = \sum_{i=1}^{N_{r}}
\left[ {\hat{r}_0(L_i) - r_0(L_i) \over \Delta r_0(L_i)} \right]^2,
\end{equation}
with $N_{\Phi}=35$ and $N_{r}=8$ the  number of data points for the LF
and the correlation lengths, respectively.

Note  that  with  this  definition  of  $\chi^2$,  we  are  implicitly
assigning  the  same weights  to  the  measurements  of $\Phi(L)$  and
$r_0(L)$.  This would  be correct if there were  no systematics in the
measurements and if  the error properties are the  same in both cases.
It is difficult to judge whether this is true in the present data, and
so our  definition of $\chi^2$  is somewhat arbitrary.   For instance,
since  $N_{\Phi}  > N_{r}$  and  since  the  errors on  $\Phi(L)$  are
typically  much smaller  than  the errors  on  $r_0(L)$, the  $\chi^2$
minimization routine gives much more  weight to fitting the LF than to
fitting  the correlation  lengths.   We will  therefore also  consider
different  relative   weights  for  the  LF   and  correlation  length
measurements,  and only  use  $\chi^2$ to  assess  the {\it  relative}
goodness-of-fit  of  different  models.    Tables~1  and  2  list  the
parameters for each  model that we discuss in  the text, together with
the corresponding values for $\chi^2(\Phi)$ and $\chi^2(r_0)$.

Before   proceeding, it   is important   to  address  the  assumptions
underlying the method. In addition  to the trivial implicit assumption
that all galaxies live inside dark matter haloes, we have assumed that
(i) $\Phi(L \vert M)$ can be  represented by a Schechter function, and
(ii) $\Phi(L   \vert M)$ is independent  of  environment; i.e.,  it is
assumed that the  galaxy properties inside a halo  of mass $M$  do not
statistically  depend on the  halo's   larger scale environment  (this
assumption only influences the  estimates of $r_0(L)$, not the fitting
of the LF). Although assumption (i) is supported  by the fact that the
LFs of  groups and clusters of galaxies  are reasonably  well fit by a
Schechter   function   (e.g.,   Muriel,   Valotto   \&   Lambas  1998;
Beijersbergen  \etal 2002; Trentham  \&   Hodgkin 2002), there is   no
strong motivation  for adopting a  Schechter form  for  haloes of much
lower  mass.   On the    other hand,  the  Schechter  function   has a
reasonable  amount of  freedom     (it is  specified  by   three  free
parameters),  and we have tested in  great detail that our results are
robust    against   different     parameterizations    of  these  free
parameters. Nevertheless, the  validity of assumption  (i) merits more
detailed   tests,  which   we   intend  to   address    in  a   future
paper. Assumption (ii) is supported by the work of Lemson \& Kauffmann
(1999) who used numerical  simulations to show  that the properties of
haloes of given mass, such  as angular momentum and concentration,  do
not   significantly  depend on    environment,   and by the   extended
Press-Schechter  formalism (e.g.  Lacey  \& Cole  1993) which predicts
that the merging history of  a halo of a given  mass is independent of
its large-scale environment.  This assumption can ultimately be tested
once better statistical samples of galaxies become available.
\begin{table*}
\begin{minipage}{\hdsize}
\caption{Model parameters for the $\Lambda$CDM ``concordance'' model.}
\begin{tabular}{lccccccccccccc}
   \hline
ID & log$M_L$ & $(M/L)_{cl}$ & $\alpha_{15}$ & log$M_1$ & log$M_2$ &
$(M/L)_0$ & $\beta$ & $\gamma_1$ & $\gamma_2$ & $\eta$ &
$\chi^2(\Phi)$ & $\chi^2(r_0)$ & $\xi_{\rm gg}$ \\
 (1) & (2) & (3) & (4) & (5) & (6) & (7) & (8) & (9) & (10) & 
(11) &(12) &(13) &(14) \\
    \hline\hline
M1  & $13.5$ & $500$ & $-1.32$ & $11.27$ & $11.73$ & $134$ & $0.77$ & $0.32$ & $0.65$ & $-
0.50$ & $41.6$ & $ 5.4$ & $+$\\
M1a & $13.5$ & $500$ & $-1.32$ & $11.34$ & $11.73$ & $133$ & $0.75$ & $0.33$ & $0.65$ & $-
0.50$ & $41.4$ & $ 5.8$ & $+$\\
M1b & $13.5$ & $500$ & $-1.32$ & $10.62$ & $11.77$ & $145$ & $0.96$ & $0.25$ & $0.66$ & $-
0.53$ &$ 90.0$ & $ 2.2$ & $+$\\
M1c & $13.5$ & $500$ & $-1.32$ & $11.88$ & $11.25$ & $ 79$ & $0.46$ & $0.52$ & $0.63$ & $-
0.65$ & $54.5$ &  $3.0$ & $+$\\
M1d & $13.5$ & $500$ & $-1.32$ & $10.42$ & $11.74$ & $102$ & $0.60$ & $0.28$ & $0.69$ & $-
0.36$ & $40.9$ & $ 1.9$ & $+$\\
 \hline
M1A & {\bf 12.5} & $500$ & $-1.32$ &  $8.64$ & $12.60$ & $163$ & $2.81$ & $0.15$ & $0.83$
& $-0.12$ & $46.1$ & $26.2$ & $+$\\
M1B & {\bf 14.5} & $500$ & $-1.32$ & $12.82$ & $12.19$ & $111$ & $0.56$ & $0.80$ & $0.87$
& $-0.31$ &$196.5$ & $43.9$ & $-$\\
M1C & $13.5$ & {\bf 250} & $-1.32$ & $10.37$ & $12.60$ & $325$ & $1.21$ & $0.05$ & $0.73$
& $-0.26$ & $54.3$ &$ 96.1$ & $+/-$\\
M1D & $13.5$ & {\bf 750} & $-1.32$ & $11.81$ & $11.49$ & $111$ & $0.57$ & $0.52$ & $0.63$
& $-0.64$ & $44.2$ & $ 3.1$ & $+$\\
M1E & $13.5$ & $500$ & {\bf -1.12} & $11.06$ & $11.93$ & $136$ & $1.07$ & $0.29$ & $0.72$
& $-0.31$ & $36.0$ & $ 4.7$ & $-$\\
M1F & $13.5$ & $500$ & {\bf -1.52} & $11.46$ & $11.53$ & $133$ & $0.64$ & $0.34$ & $0.60$
& $-0.70$ & $49.2$ & $ 6.6$ & $-$\\
 \hline
\end{tabular}
\medskip

Column~(1) lists the  ID by which we refer to each  model in the text.
Columns~(2)  to~(4) list model  parameters that  are kept  fixed; here
$M_L$ is  defined such that $\wLstar(M_L)=L^{*}$,  $(M/L)_{\rm cl}$ is
the mass-to-light  ratio of haloes  with $M \geq 10^{14}  \msunh$, and
$\alpha_{15}$ is the faint-end slope  of the conditional LF for haloes
with $M=10^{15}  h^{-1} \Msun$. Columns~(5) to~(11)  list the best-fit
model parameters, of  which $M_2$ and $\gamma_1$ are  set by $M_L$ and
$(M/L)_{cl}$, respectively.  Columns~(12) and  (13) list the values of
$\chi^2(\Phi)$    and   $\chi^2(r_0)$    of   the    best-fit   model,
respectively. Finally, column~(14) indicates  whether the shape of the
galaxy correlation functions are  in good ($+$), reasonable ($+/-$) or
poor ($-$)  agreement with the data.  Masses  and mass-to-light ratios
are in $h^{-1} \Msun$ and $h \MLsun$, respectively.

\end{minipage}
\end{table*}

\subsection{The concordance model}
\label{sec:concordance}

\begin{figure*}
\centerline{\psfig{figure=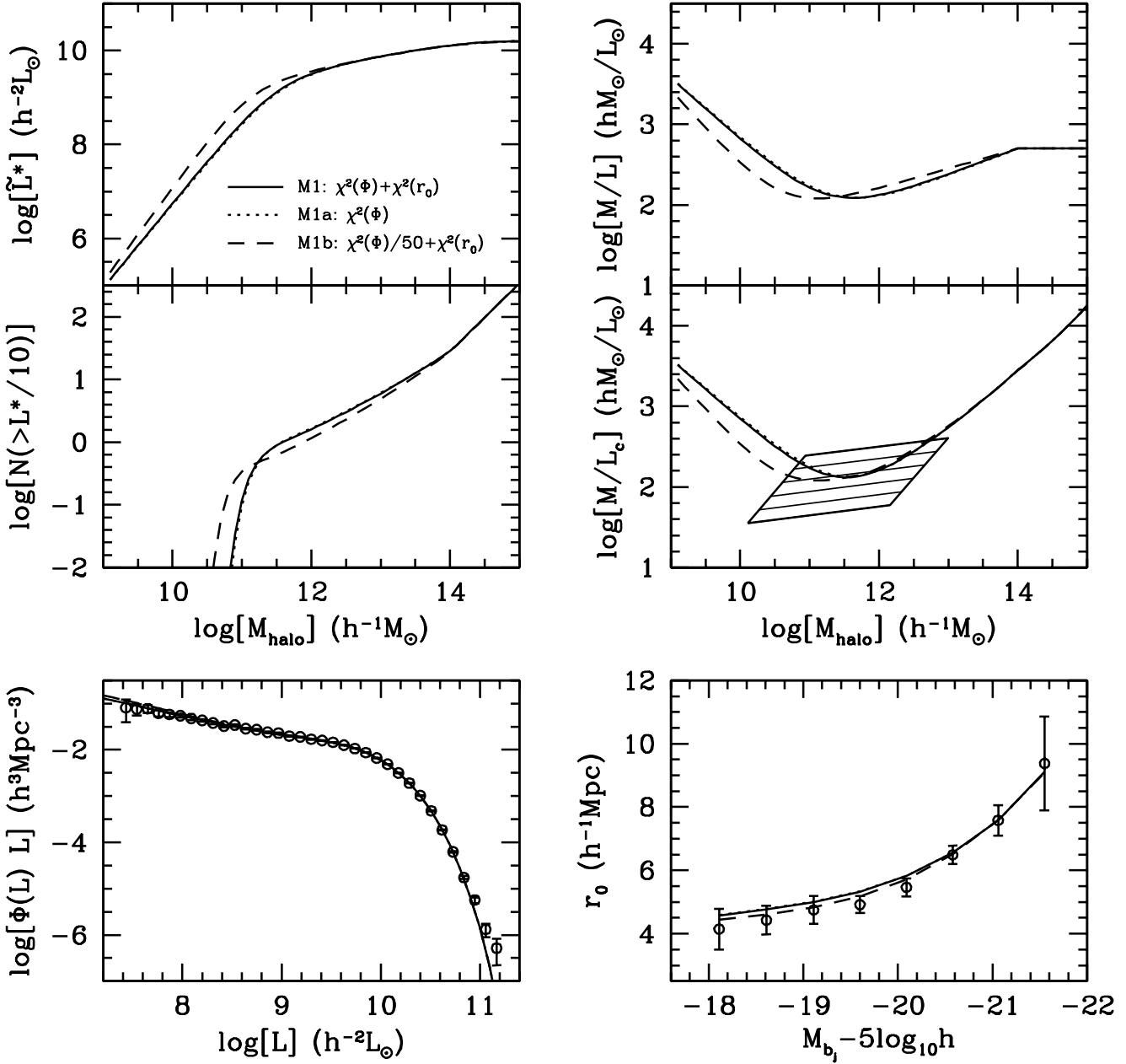,width=\hdsize}}
\caption{Results for the fiducial  model  M1 (solid lines).  The   two
bottom  panels compare the  model   predictions with the  observed  LF
$\Phi(L)$ (left) and   correlation  lengths  $r_0(L)$  (right).   Open
circles with errorbars   correspond to the  2dFGRS  data  from Norberg
\etal  (2001a,b). The four top panels  show, clockwise  from the upper
left panel,  the  characteristic  luminosity,  ${\tilde L}_\star$, the
ratio between halo mass and total  luminosity $M/L$, the ratio between
halo mass and  luminosity of  the `central'  galaxy $M/L_c$,  and  the
number of  galaxies with $L>L_{*}/10$, all as   function of halo mass.
The dotted and dashed curves  show the results obtained by  minimizing
$\chi^2=\chi^2(\Phi)$    (Model  M1a)   and $\chi^2=\chi^2(\Phi)/50  +
\chi^2(r_0)$ (Model M1b), respectively.}
\label{fig:concmodel}
\end{figure*}

We first consider the standard $\Lambda$CDM model with $\Omega_0=0.3$,
$\Omega_{\Lambda} = 0.7$,  $h = 0.7$, $\Gamma =\Omega_0  \, h = 0.21$,
and $\sigma_8=0.9$.  These cosmological parameters are consistent with
a wide range  of observations, and so this  cosmology has become known
as the ``concordance'' model (e.g., Perlmutter \etal 1999; Riess \etal
1998, 2001; Melchiorri \etal 2000; de Bernardis \etal 2002).

The solid  curves in Figure~\ref{fig:concmodel}  show  the results for
the best fit model (M1) for this  concordance cosmology.  Hereafter we
refer to this model as our `fiducial' model,  and compare other models
with respect to it.  The bottom left panel  compares the model LF with
that of  the 2dFGRS (open circles with  errorbars).   The agreement is
excellent,  which is also apparent from   the fact that $\chi^2(\Phi)$
divided  by the number  of degrees of freedom is  very close to unity.
The agreement with the observed $r_0(L)$ (shown in the bottom panel on
the right)  is also acceptable, although   the model predicts slightly
too large correlation lengths for galaxies with $M_{b_j} - 5 {\rm log}
h \gta -20$.  Note that the  functional form of $r_0(L)$ is remarkably
well reproduced by the model.

The  best-fit  value  of  $\eta$  is  $-0.50$,  which  indicates  that
$\walpha$ decreases with increasing  mass.  Larger values of $\walpha$
imply  a  LF that  is  more peaked  at  around  $\wLstar$, i.e.,  with
relatively fewer galaxies  with $L < \wLstar$.  Note  that a faint-end
slope that  decreases with increasing  mass is at  least qualitatively
consistent  with  observational  results  (e.g., Trentham  \&  Hodgkin
2002).   The top  left panel  of Figure~\ref{fig:concmodel}  plots the
characteristic luminosity, $\wLstar$, as function of mass.  A best-fit
value for  $\beta$ of $0.77$  implies that $\wLstar  \propto M^{1.77}$
for $M \ll  M_1 \simeq 1.9 \times 10^{11} h^{-1}  \Msun$. The slope of
$\wLstar(M)$  for haloes  with $M  \gg  M_1$ is  mainly controlled  by
$\gamma_2$,  and is  strongly  constrained by  the  form of  $r_0(L)$;
Larger (smaller) values of $\gamma_2$ cause $r_0(L)$ to increases less
(more) rapid  with increasing luminosity.  Note that for  our best-fit
value of  $\gamma_2=0.65$ the value  of $\wLstar$ only increases  by a
factor  $\sim 5$ from  $M=10^{12} h^{-1}  \Msun$ to  $M=10^{15} h^{-1}
\Msun$.

The middle panel on the  left plots $N(>L^{*}/10)$, the average number
of galaxies  brighter than one-tenth of  the characteristic luminosity
$L^{*}$ of the 2dFGRS.  For  $M \gta 10^{12} h^{-1} \Msun$ this number
increases roughly linearly with mass, while below $\sim 10^{11} h^{-1}
\Msun$ one basically no longer finds galaxies as bright as $L^{*}/10$.
The top and  middle panels on the right  plot the mass-to-light ratios
$M/L$  and  $M/L_c$, respectively.   Here  $L$  is  the average  total
luminosity   of    galaxies   in   a    halo   of   mass    $M$   (see
equation~[\ref{LofM}]),   and  $L_c$   corresponds   to  the   average
luminosity of  the `central galaxy'  (equation~[\ref{Lcentral}]).  The
mass-to-light $M/L$ reaches a minimum  value of $\sim 125 h \MLsun$ at
$M  \sim 10^{11.5}h^{-1}\msun$.   For  haloes with  masses smaller  or
larger   than  $\sim   10^{11.5}h^{-1}\msun$,  the   galaxy  formation
efficiency  is reduced,  consistent with  the fact  that  feedback can
reduce the  star formation efficiency  in small haloes,  while cooling
becomes less  effective in  more massive haloes  (e.g., White  \& Rees
1978;  Dekel \& Silk  1986; van  den Bosch  2002; Benson  \etal 2000).
Note that, by construction,  the mass-to-light ratio stays constant at
$M/L=500   h    \MLsun$   for    $M   \geq   10^{14}    \msunh$   (see
Section~\ref{sec:condLF}).

\begin{figure*}
\centerline{\psfig{figure=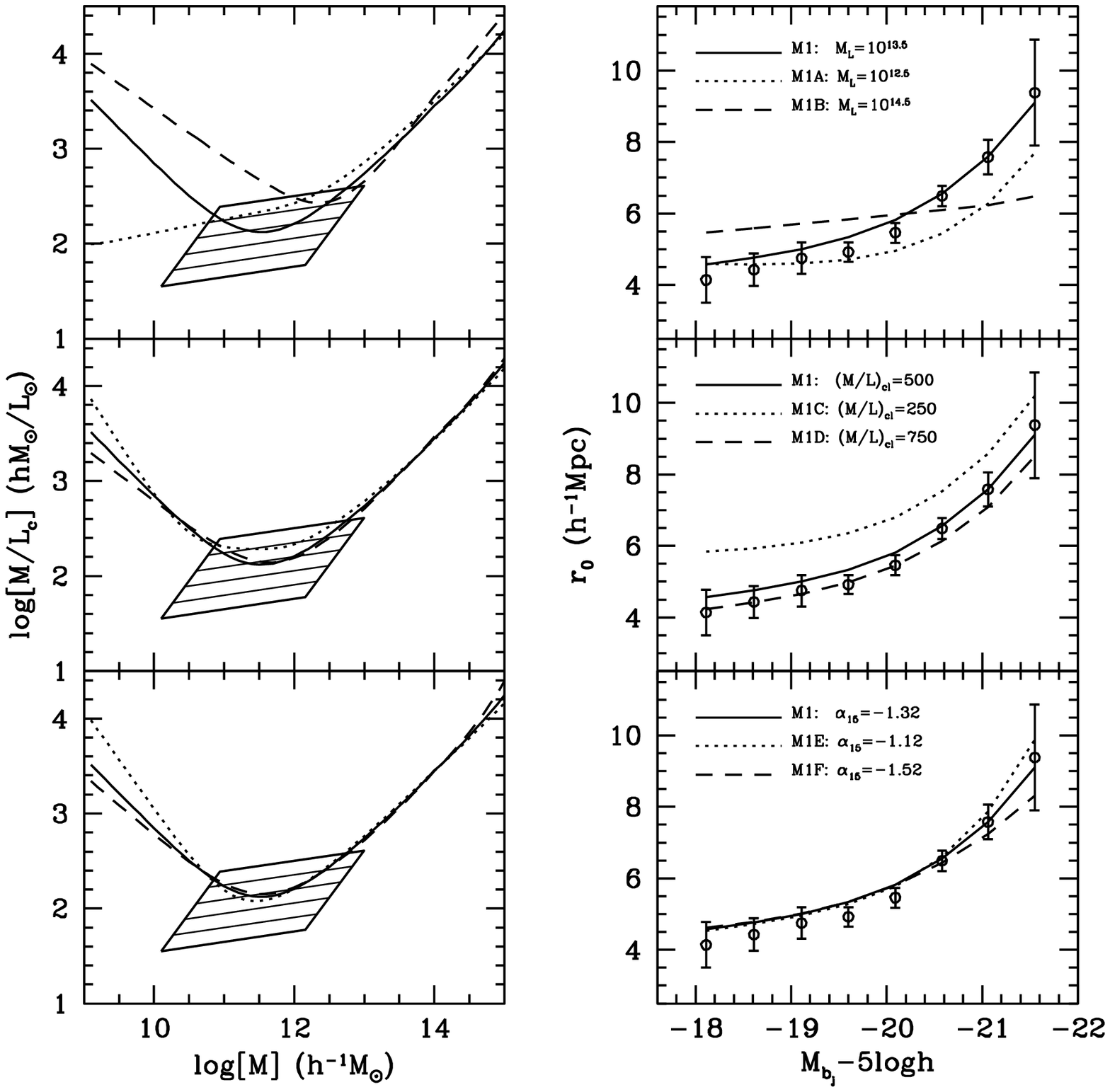,width=\hdsize}}
\caption{Results of models in  which  one model parameter  is  changed
relative to the fiducial choice.  Left panels show $M/L_c$ as function
of halo mass (parallelogram  indicates the TF constraints),  while the
right panels  show the  correlation length   as  function of  absolute
magnitude.    Open    circles with   errorbars    correspond   to  the
observational  data from the    2dFGRS.   Solid curves in each   panel
correspond to  the fiducial  model  M1.  Top panels  show results  for
models with $M_L=10^{12.5}    \msunh$ (Model M1A, dotted  curves)  and
$M_1=10^{14.5}h^{-1} \msun$ (Model M1B, dashed curves).  Middle panels
show results for  models with $(M/L)_{\rm  cl} = 250 h \MLsun$  (Model
M1C, dotted curves)  and $(M/L)_{\rm cl} =  750 h \MLsun$  (Model M1D,
dashed curves). Bottom panels  show results for $\alpha_{15} =  -1.12$
(Model M1E, dotted  curves) and $\alpha_{15}=-1.52$ (Model M1F, dashed
curves). See text for a detailed discussion.}
\label{fig:varmodel}
\end{figure*}

For low  mass haloes,  the  ratio  $M/L_c$  has properties similar  to
$M/L$.   This owes to the  fact that, by   construction, each of these
haloes hosts only    one  dominant (`central')   galaxy.  For  massive
systems, the faint-end   slope of the   conditional LF is  steeper and
${\cal   N}^{*}  >  1$, causing   $M/L_c   >  M/L$.  The parallelogram
indicates   the  TF constraints for   six  different values of $V_{\rm
max}/V_{\rm vir}$ from  $1.0$ (upper line)  to $2.0$ in steps of $0.2$
(see Figure~\ref{fig:tulfis}).  In order for model M1 to be consistent
with the zero-point of the  TF relation, the ratio $V_{\rm max}/V_{\rm
vir}$ must be smaller than $1.4$,  which is only marginally consistent
with CDM predictions, especially  when   the boosting effect of    the
baryonic matter   is    taken  into    account  (see   discussion   in
Section~\ref{sec:tf}).  In addition, the fact that $M/L_c(M)$ is quite
strongly `curved'  over the mass range  of the  TF constraints implies
that $V_{\rm max}/V_{\rm  vir}$ needs to vary  significantly, and in a
fine-tuned   fashion, with   mass  in  order  to   yield  a linear  TF
relation\footnote{Note  that the discrepancy for  haloes  with $M \gta
10^{12} h^{-1} \Msun$ may not be a too  serious problem, because their
central galaxies  may  be predominantly  ellipticals for which  the TF
relation does not apply.}.  Thus although model M1 yields a reasonable
fit  to the LF   and $r_0(L)$, it  predicts  values for $\langle M/L_c
\rangle$ that are slightly too high.

As pointed out before, by virtue of the definition of $\chi^2$ used in
the minimization routine, much more weight is  given to fitting the LF
than to fitting the correlation  lengths.  Consequently, the fits  are
expected  to   be controlled  mainly   by  the   LF   constraints.  To
demonstrate  this,  the dotted curves  in  Figure~\ref{fig:concmodel},
corresponding  to model M1a,  show  the results obtained when  fitting
only the LF, i.e., by setting $\chi^2  = \chi^2(\Phi)$.  The fact that
these curves are all very similar to our fiducial model (solid curves)
confirms  the above  assertion.   In addition,  the  dashed  curves in
Figure~\ref{fig:concmodel},  corresponding  to   model   M1b, show the
results obtained by setting $\chi^2  = \chi^2(\Phi)/50 + \chi^2(r_0)$,
i.e.  by reducing  the  weights of the   $\Phi(L)$ measurements  by  a
factor  $50$.   Not surprisingly,   this  model fits  the  correlation
lengths better  but fits the LF  slightly worse, at least according to
the value of   $\chi^2(\Phi)$.  Whether or  not the  fit is acceptable
depends on possible systematic errors  in the observed $\Phi(L)$ which
may be larger than the statistical errors used here.  We return to the
issue of systematic errors in Section~\ref{sec:LFuncertain}.

\subsection{Less restrictive models}
\label{sec:moduncertain}

The  results  presented  above  suggest  that,  for  the  cosmological
concordance  model, the  conditional luminosity  distribution obtained
from the observed LF and  the observed clustering strength as function
of  luminosity over-predicts  the mass-to-light  ratios implied  by the
Tully-Fisher relation.   However, this  assertion is only  correct for
our restricted set of parameterized models for $\Phi(L \vert M)$.  The
more general model for our  conditional LF has nine, rather than five,
free  parameters.   Although  we   have  argued  that  four  of  these
parameters  can   be  fixed   based  on  physical   and  observational
constraints, we  now address the  impact these `restrictions'  have on
our results.

In our fiducial model we  fix the mass $M_2$ by setting $M_L=10^{13.5}
h^{-1}\msun$,  where   $M_L$  is  defined  so   that  $\wLstar(M_L)  =
L^{*}$. This  choice is  motivated by the  fact that $L^{*}$  does not
seem  to  vary much  over  the range  $10^{13}  {\msunh}  \lta M  \lta
10^{15}{\msunh}$.  To  illustrate the impact that  this assumption has
on our results, the upper panels of Figure~\ref{fig:varmodel} plot the
results for models M1A and M1B where we set $M_L=10^{12.5} \msunh$ and
$M_L  =   10^{14.5}  \msunh$,  respectively.    Note  that  increasing
(decreasing)  the value  of  $M_L$ mainly  causes  bright galaxies  to
reside in  less (more)  massive haloes.  As  one can see,  the results
depend quite sensitively on the value of $M_L$. For $M_L \gg 10^{13.5}
\msunh$  the predicted  $r_0 (L)$  only shows  a very  mild luminosity
dependence, in  clear disagreement with  the data.  In  addition, this
model yields values  of $\chi^2(\Phi)$ that are too  high and predicts
galaxy-galaxy  correlation functions  that deviate  strongly  from the
power-law  form  for  $M_{b_j}  -  5  {\rm log}  h  \lta  -21.0$  (see
Section~\ref{sec:corrshape}  below).   Model  M1A,  for which  $M_L  =
10^{12.5}  \msunh$, also  fails  to match  the  observed $r_0(L)$  for
bright  galaxies. Both  models predict  mass-to-light ratios  that are
clearly inconsistent  with the TF constraints.   We therefore conclude
that models  with $10^{13}  {\msunh} \lta M_L  \lta 10^{13.5}{\msunh}$
are the  most successful, and in  what follows we keep  $M_L$ fixed at
$10^{13.5} \msunh$.

In the  fiducial model, we tune  $\gamma_1$ so that  $M/L = (M/L)_{\rm
cl} = 500 h  \MLsun$ at $M = 10^{14} h^{-1} \Msun$,  and we keep $M/L$
fixed at the value of  $(M/L)_{\rm cl}$ for more massive systems.  The
middle  two panels  of Figure~\ref{fig:varmodel}  show the  effects of
changing $(M/L)_{\rm  cl}$ to  $250 h \MLsun$  (model M1C) and  $750 h
\MLsun$  (model  M1D). These  values  are  still  consistent with  the
observational constraints  of Fukugita \etal (1998) at  the $3 \sigma$
level.   Increasing (decreasing)  $(M/L)_{\rm cl}$  causes  a decrease
(increase) of  the number  of galaxies in  massive haloes.   Since the
clustering  strength  of  haloes  increases  with  mass,  an  increase
(decrease)  of $(M/L)_{\rm  cl}$ therefore  results in  lower (higher)
values of  $r_0(L)$. For  $(M/L)_{\rm cl} <  500 h \MLsun$  the models
predicts too large correlation lengths as well as mass-to-light ratios
that are  inconsistent with the  TF constraints.  Model  M1D, however,
predicts  a  $\hat{r}_0(L)$  and  $\hat{\Phi}(L)$  that  are  both  in
agreement  with  the data.   Over  the mass  range  probed  by the  TF
constraints, the mass-to-light ratios are almost identical to those of
the  fiducial  model;  they  are  slightly  too  high  and  reveal  an
uncomfortably large variation over this mass range.  Since $(M/L)_{\rm
cl}  =  750   h  \MLsun$  is  only  marginally   consistent  with  the
observational  constraints  of  Fukugita  \etal  (1998),  we  restrict
ourselves to $(M/L)_{\rm cl} = 500 h \MLsun$ in what follows.

The  lower panels  of Figure~\ref{fig:varmodel}  compare  the fiducial
model  (for which $\alpha_{15}  = -1.32$)  to models  M1E and  M1F for
which   $\alpha_{15}    =   -1.12$   and    $\alpha_{15}   =   -1.52$,
respectively. The choice  of $\alpha_{15}$ has only a  small impact on
the  results.  A  change in  $\alpha_{15}$ is  largely  compensated by
changes  in $M_1$,  $\beta$  and $\eta$,  but  does not  significantly
improve or worsen the  agreement with the data. However, $\alpha_{15}$
does    impact   on    the   shape    of   $\xi_{\rm    gg}(r)$   (see
Section~\ref{sec:corrshape}  below)  with $\alpha_{15}=-1.32$  clearly
giving the  best results.  We  therefore restrict ourselves  to models
with $\alpha_{15}=-1.32$ in what follows.

Finally we  investigated the effect  of changing $M_{\rm  min}$, i.e.,
the mass scale below which $\Phi(L \vert M) = 0$. We find that as long
as $M_{\rm min} \lta 10^{10}  \msunh$ our results do not significantly
depend on  $M_{\rm min}$, and  our fiducial value of  $10^{9} \msunh$,
which is motivated by re-ionization considerations, therefore does not
influence any of the results presented here.

In addition  to   varying some of  the  parameters  kept fixed in  our
fiducial models,  we   also  experimented with  completely   different
parameterized       forms       of  $\wLstar$,     $\walpha$,      and
$\tilde{\Phi}^{*}$. Neither  of these alternative models, however, was
able to  yield results in better agreement  with the data.  Therefore,
we conclude   that  the $\Lambda$CDM  concordance model   can be  made
consistent with both  the observed LF and  the correlation lengths  as
function of   luminosity.  The largest   problem  for the cosmological
concordance model is to fit the zero-point of the  TF relation. In all
the  models   presented   here, the  mass-to-light   ratios  are  only
consistent with the TF constraints if  $V_{\rm max} / V_{\rm vir} \lta
1.4$, which  is hard  to reconcile with  typical CDM  predictions (see
discussion in Section~\ref{sec:tf}).  Probably equally problematic, is
the  fact  that the  models  predict  mass-to-light ratios  that  vary
strongly over the mass range probed by the TF constraints.
\begin{figure*}
\centerline{\psfig{figure=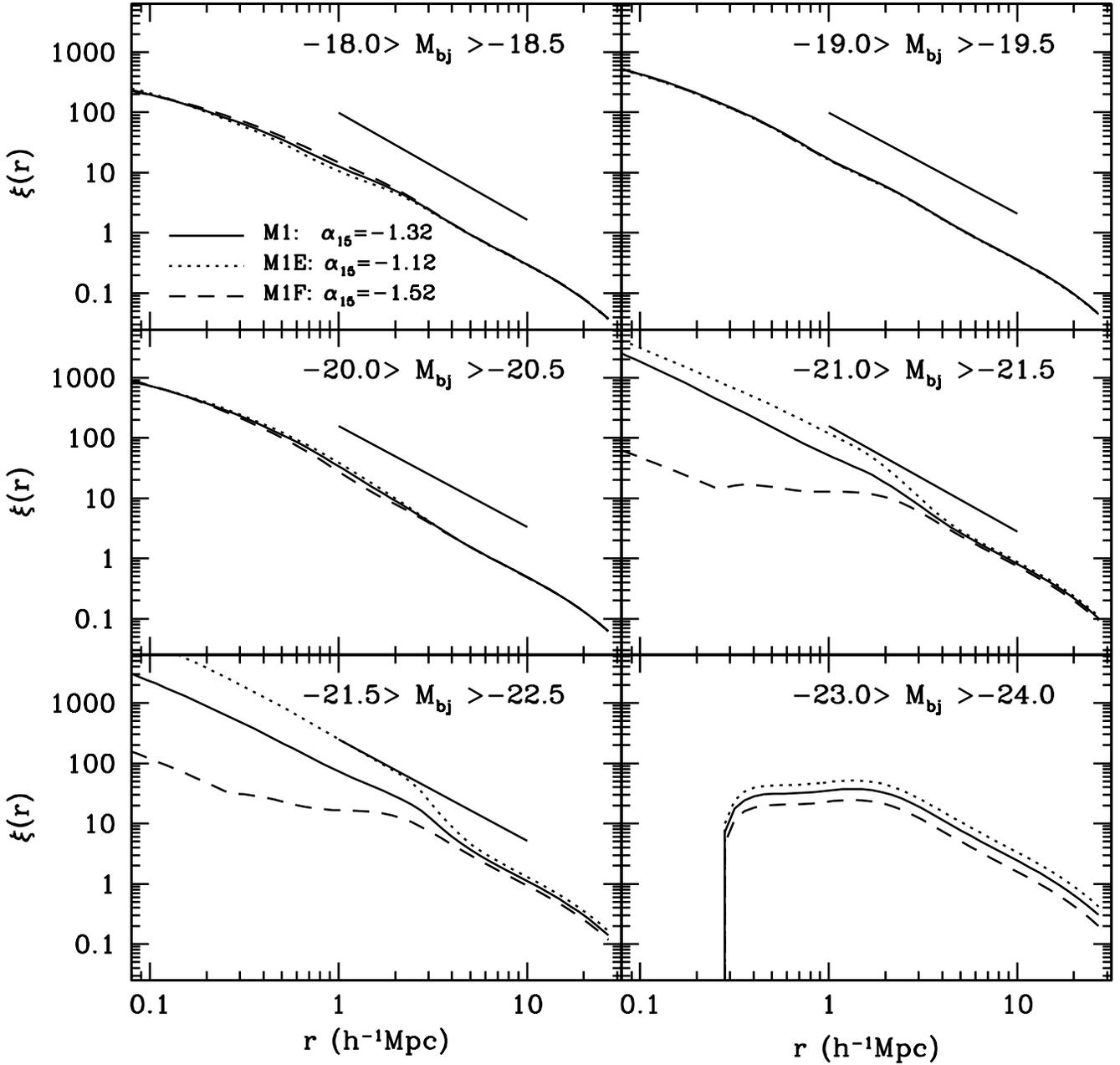,width=\hdsize}}
\caption{The predicted  two-point correlation functions of galaxies in
different absolute magnitude bins.  Results are shown for the fiducial
model    (Model   M1;    solid     curves),  and  for    models    M1E
($\alpha_{15}=-1.12$;  dotted curves)  and   M1F ($\alpha_{15}=-1.52$;
dashed curves).  The short  solid line-segments indicate the slopes of
the best-fit power-laws obtained  by Norberg \etal (2001a) for  2dFGRS
galaxies in the same magnitude bins.  No such line-segment is shown in
the lower  right panel, since  no 2dFGRS  data  is available  for this
magnitude bin.  Note that the   correlation functions of the  fiducial
model resemble power-laws over  the entire radial range plotted,  with
slopes that are very  similar to those observed.   An exception is the
magnitude    bin in the   lower  right panel,   which  reveals a clear
deviation from a power law at small  separations.  Models M1E and M1F,
on the  other hand, show significant deviations  from a pure power law
behavior for galaxies with $-21 > M_{b_j} - 5 {\rm log} h > -22.5$.}
\label{fig:shape}
\end{figure*}

\subsection{The shape of the correlation function}
\label{sec:corrshape}

It is well known that the observed correlation function of galaxies is
well fit  by  a single  power law  $\xi_{\rm gg}(r)=(r/r_0)^{-\gamma}$
over a wide range  of radii (e.g.   Davis \& Peebles 1983; Baugh 1996;
Jing, Mo \& B\"orner 1998;  Zehavi \etal 2002).  Norberg \etal (2001a)
have  recently shown  that even  when $\xi_{\rm   gg}(r)$ is split  in
several luminosity bins, each individual correlation function is still
well  fit by  a single  power  law  with  $\gamma  \simeq 1.7$.  These
observations yield  extra  constraints on  the  conditional luminosity
function.  Note that the mass correlation function of the CDM model in
consideration  is {\it not} well  described by a   pure power law (see
Fig.\,\ref{fig:xi}).  The power-law  behavior of $\xi_{\rm gg}(r)$ for
individual  luminosity  bins therefore results   from some non-trivial
conspiracy between the halo bias on large scales (the 2-halo term) and
the distribution of galaxies within individual dark matter haloes (the
1-halo term) on small scales.  This  property of the galaxy occupation
model has been noticed in some early investigations  (e.g. Jing, Mo \&
B\"orner 1998; Jing,  B\"orner  \& Suto  2002;  Kauffmann \etal  1999;
Benson  \etal 2000; Berlind \& Weinberg  2002).

Figure~\ref{fig:shape} plots  the galaxy-galaxy  correlation functions
of model M1  (solid  lines) for  five   luminosity bins also used   by
Norberg \etal (2001a).  In addition,  we also plot predictions for the
magnitude bin   $-24 < M_{b_j}    - 5 {\rm  log} h   <  -23$. For  all
sub-samples with $M_{b_j}  - 5 {\rm  log} h > -22.5$,  the correlation
functions have roughly a power-law form over the  entire range from $r
\simeq 0.1 h^{-1} \Mpc$ to $r \simeq 20 h^{-1} \Mpc$, and with a slope
close to that measured from the 2dFGRS (indicated in each panel).  For
galaxies with $-23.0 > M_{b_j} - 5 {\rm log} h  > -24.0$, however, the
correlation  function is strongly suppressed at  $r\la 2\mpch$.  These
bright galaxies correspond to  the central galaxies of  rich clusters,
and  so their   correlation   function  contains only   a   negligible
contribution from the 1-halo term.   Unfortunately, the 2dFGRS results
of  Norberg \etal (2001a)  only considered galaxies  with $M_{b_j} - 5
{\rm log} h >  -22.5$, so that we cannot  compare this prediction with
data. Figure~\ref{fig:shape} also plots the predictions for models M1E
($\alpha_{15}=-1.12$)    and  M1F  ($\alpha_{15}=-1.52$).   Both these
models show  significant deviations from  a single power-law  form (at
least for the brighter  galaxies), demonstrating that the actual shape
of the   correlation  function can  be used   to  discriminate between
different models  for the conditional  LF.  It is remarkable  that the
model  with  $\alpha_{15}=-1.32$,  which is    the value  obtained  by
Beijersbergen  \etal  (2002) for the  Coma  cluster,  yields the  best
result.

Although  the shapes  of the  two-point correlation  functions  of our
fiducial model are in good  agreement with the 2DFGRS, this success is
model-dependent. In particular, we  have assumed that the distribution
of galaxies  in individual haloes follows the  density distribution of
dark  matter, and we  have made  simplified assumptions  regarding the
second  moment of  the distribution  of the  halo  occupation numbers.
Indeed,  as shown  in Berlind  \& Weinberg  (2002), the  shape  of the
correlation function  on small scales may depend  sensitively on these
assumptions, and a  good power law is not  always obtained. Therefore,
in judging the success of our  models we will not give too much weight
to the  actual shape of  $\xi_{\rm gg}(r)$. For completeness,  we {\it
do}  indicate in  Tables~1  and~2 whether  the correlation  functions,
obtained under the  assumptions of our model for  the 1-halo term, are
in good  ($+$), reasonable  ($+/-$) or poor  ($-$) agreement  with the
data.

\begin{figure*}
\centerline{\psfig{figure=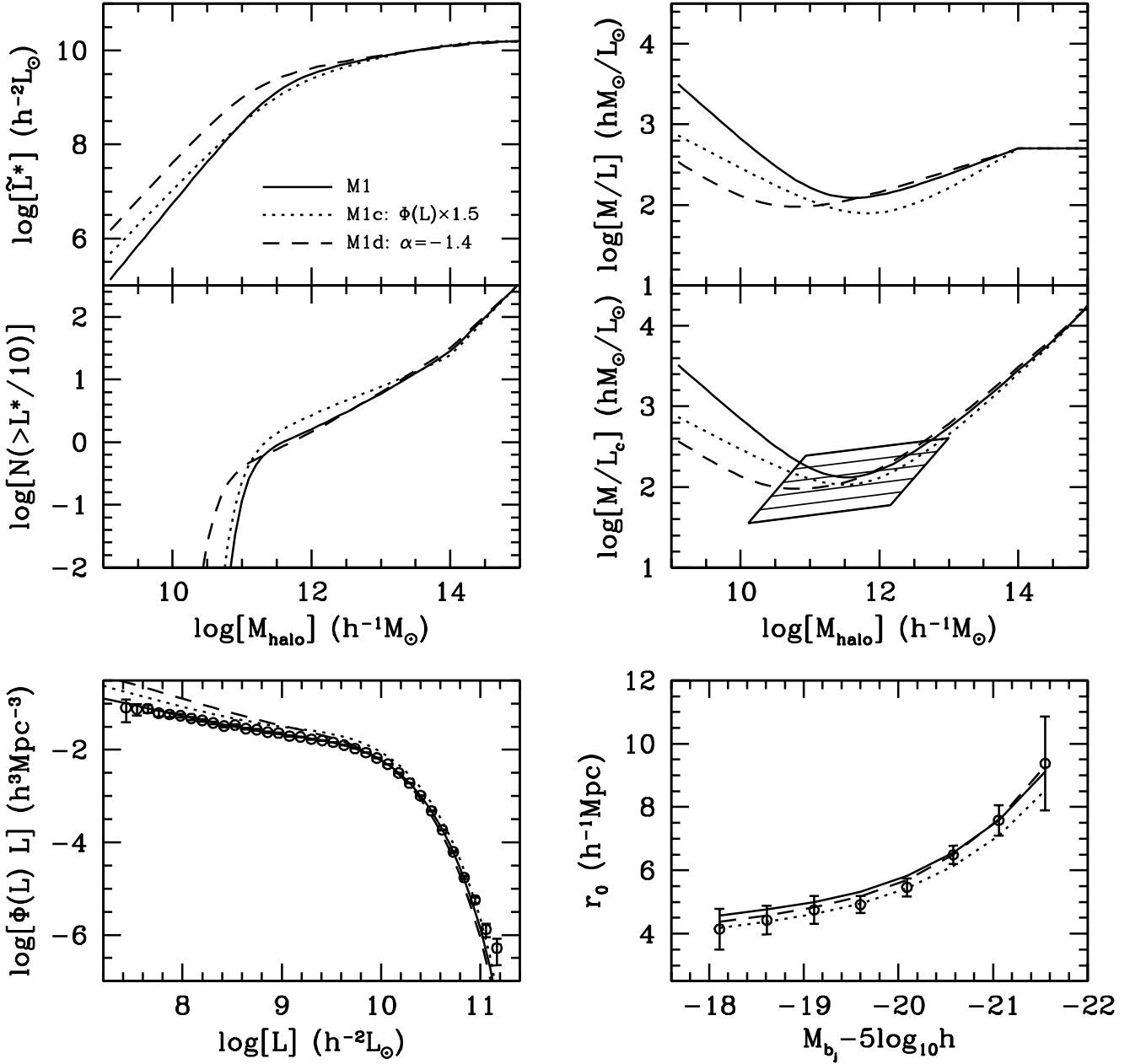,width=\hdsize}}
\caption{Same   as   Figure~\ref{fig:concmodel} except  that   here we
compare the fiducial model M1 with models in  which we assume that the
true LF differs from the 2dFGRS LF. In particular, we show results for
models in which we assume that the true  number density of galaxies is
$1.5$ times as  high as for  the 2dFGRS LF, independent  of luminosity
(Model M1c; dotted curves), and in which we assume that the true faint
end slope is $\alpha =  -1.4$ in stead of  $-1.21$ (Model M1d;  dashed
curves). Note that both models results  in lower values of $M/L_c$ (in
better agreement with the TF  constraints) and in smaller  correlation
lengths (in better agreement with the 2dFGRS data).}
\label{fig:errors}
\end{figure*}

\section{Systematic errors in the observational constraints}
\label{sec:obsuncertain}

As we have shown above, none of the $\Lambda$CDM concordance models is
fully consistent with all  observational constraints.  Here we address
whether  this  discrepancy  might  result  from  uncertainties  and/or
systematic errors in the observational constraints.

\subsection{The luminosity function}
\label{sec:LFuncertain}

The  determination of  an accurate LF   is  non-trivial.  The accuracy
depends,  amongst  others, on the size  of  the  redshift  survey (the
larger   the survey, the smaller  the  systematic  errors due to large
scale structure), on surface  brightness completeness effects, on  the
accuracy  of star-galaxy  separation,  and  on  the  accuracy   of the
photometric calibration.  In addition, corrections have to be made for
band-shifting (the $k$-correction) and for evolutionary effects of the
galaxy   population  with  redshift,   each   of  which  has   its own
uncertainties.  Our main motivations for using the  2dFGRS LF are that
(i) it is  the largest extant  redshift survey, (ii) its quoted errors
have  contributions  from all the effects   mentioned above, and, most
importantly, (iii) the differential galaxy clustering properties, used
as additional constraints  on the model,   have been derived  from the
same 2dFGRS data set.

Despite these  advantages over other LFs available  in the literature,
one  cannot  rule  out  that  some amount  of  systematic  errors  are
present. For instance,  Blanton \etal (2001) claim that  the 2dFGRS LF
(initially presented by Folkes \etal 2001) is inconsistent with the LF
of the SDSS commissioning  data. Norberg \etal (2001b), however, claim
that  after taking  account of  zero-point photometric  offsets, after
correcting for galaxy evolution,  and after making corrections for the
proper  conversion between SDSS  and 2dFGRS  magnitudes, both  LFs are
actually in good agreement.

Despite this apparent concordance, it is worthwhile to investigate how
the presence of certain systematic errors impacts on our results.  One
of the  most likely sources  for systematic errors  is incompleteness.
Note that the incompleteness corrections applied by Norberg \etal were
based on a  comparison with the SDSS early-data  release, which itself
is  not expected to  be 100  percent complete.   If indeed  the 2dFGRS
suffers from incompleteness effect, the true luminosity densities will
be higher, resulting in lower mass-to-light ratios for individual dark
matter haloes.   This is exactly  the effect that might  alleviate the
problems with the TF constraints outlined above.  We therefore address
the impact incompleteness may have  on our results using two different
approaches.

First we  consider an overall,  luminosity independent incompleteness,
by assuming  that $\Phi^{*}$ of the  2dFGRS is a factor  1.5 too small
(Model M1c).   Note that this  is roughly similar to  the disagreement
between  the SDSS  and 2dFGRS  LFs  claimed by  Blanton \etal  (2001).
Secondly  we  consider   the  possibility  of  a  luminosity-dependent
incompleteness by assuming  that the true value of  $\alpha$ is $-1.4$
rather   than   $-1.21$   (Model   M1d).    Results   are   shown   in
Figure~\ref{fig:errors}.   Not  surprisingly,  both models  result  in
mass-to-light ratios  in better  agreement with the  TF relation.
Thus, the  $\Lambda$CDM concordance model can be  made consistent with
all observational  constraints considered here  if the 2dFGRS  is only
about $70$ percent complete  at all luminosities. Given the discussion
in Norberg \etal (2001b) such extreme incompleteness effects, however,
seem unlikely.

The incompleteness effects discussed here have exactly the same effect
as the    presence of dark   CDM haloes   (i.e.,  haloes  in  which no
observable galaxies formed)\footnote{Note that we have already assumed
that haloes with $M < M_{\rm min}$  are dark, which we associated with
re-ionization effects.}. In this case, model M1c  is to be interpreted
as a model in which one third of all haloes (independent of mass) host
no observable galaxies (i.e., $\Phi(L \vert  M)=0$), while in the case
of model M1d the fraction of dark CDM haloes increases with decreasing
halo mass.  The possibility of  dark CDM haloes was recently discussed
by Verde,  Oh  \& Jimenez  (2002), who argued   that haloes with large
angular momenta might  result in disks of  such low surface brightness
that  star formation  is   never triggered   (see also Jimenez   \etal
1997). As shown  by Jimenez, Verde \&  Oh (2002) a large abundance  of
dark  CDM   haloes is consistent   with the  rotation   curves of disk
galaxies.  Based  on our discussion  above, we therefore conclude that
the disagreement  of the $\Lambda$CDM  concordance  model with  the TF
constraints may  be alleviated if a  relatively large fraction of dark
matter haloes never managed to form observable galaxies.

\subsection{The correlation lengths}
\label{sec:corruncertain}

As  we have  argued  above, the  differential  clustering strength  of
galaxies  as  function  of  luminosity provides  important  additional
constraints that  allow us to  obtain a unique best-fit  $\Phi(L \vert
M)$. In particular, the value of $\gamma_2$ is strongly constrained by
the strong increase in correlation length at the bright end.

In  this paper,  we used  the results  from the  2dFGRS,  presented by
Norberg  \etal   (2001a),  for   the  reasons  already   mentioned  in
Section~\ref{sec:LFuncertain}.  A comparison  of the $r_0(L)$ obtained
from the 2dFGRS  with those obtained from a  variety of other surveys,
in particular the  Southern Sky Redshift Survey 2  (Benoist \etal 1996;
Willmer \etal  1998), the  Stromlo/APM redshift survey  (Loveday \etal
1995), and the ESO Slice  Project (Guzzo \etal 2000), shows an overall
agreement  that  $r_0(L)$ increases  with  increasing luminosity  (see
Figure~3  in  Norberg  \etal  2001a).   Quantitatively,  however,  the
disagreement  is relatively  large compared  to the  quoted errorbars.
Undoubtedly,  to a  large extent  this owes  to  sampling fluctuations
associated with  the relatively  small size of  some of  these surveys
(e.g., Benson  \etal 2001). Since the  2dFGRS is currently  by far the
largest  redshift survey  for which  differential  clustering measures
have been obtained, we adhere to  these data. In the near future, when
the full 2dFGRS  and SDSS results are available,  the true accuracy of
the $r_0(L)$ measurements used here can be addressed in more detail.
\begin{table*}
\begin{minipage}{\hdsize}
\caption{Model parameters for different cosmologies.}
\begin{tabular}{lcccccccccccccccc}
    \hline
ID & $\Omega_0$ & $\Omega_{\Lambda}$ & $\sigma_8$ & $n$ &
$R_f$ & $m_{X}$ & log$M_1$ & log$M_2$ &
$(M/L)_0$ & $\beta$ & $\gamma_1$ & $\gamma_2$ & $\eta$ &
$\chi^2(\Phi)$ & $\chi^2(r_0)$ & $\xi_{\rm gg}$\\
 (1) & (2) & (3) & (4) & (5) & (6) & (7) & (8) & (9) & (10) & 
(11) &(12) &(13) &(14) &(15) &(16) &(17)\\
    \hline\hline
M2 & {\bf 0.2} & $0.8$ & {\bf 1.06} & $1.0$ & $-$ & $-$ & $11.25$ & $11.18$ &  $72$ & $0.6
0$ & $0.42$ & $0.64$ & $-0.73$ & $60.5$ &$ 72.4$ & $+$ \\
M3 & {\bf 0.3} & $0.7$ & {\bf 0.86} & $1.0$ & $-$ & $-$ & $11.42$ & $11.69$ & $132$ & $0.7
2$ & $0.34$ & $0.65$ & $-0.52$ & $41.6$ & $ 2.5$ & $+$ \\
M4 & {\bf 0.4} & $0.6$ & {\bf 0.74} & $1.0$ & $-$ & $-$ & $11.10$ & $12.16$ & $226$ & $1.0
2$ & $0.22$ & $0.66$ & $-0.40$ & $45.8$ & $12.9$ & $+$ \\
M5 & {\bf 0.5} & $0.5$ & {\bf 0.66} & $1.0$ & $-$ & $-$ & $10.92$ & $12.58$ & $369$ & $1.4
0$ & $0.14$ & $0.69$ & $-0.31$ & $45.9$ & $42.9$ & $+$ \\
M6 & {\bf 1.0} & $0.0$ & {\bf 0.46} & $1.0$ & $-$ & $-$ & $ 8.74$ & $14.34$ &$5674$ & $2.8
9$ &$-0.14$ & $0.83$ & $-0.09$ &$305.4$ &$187.0$ & $+$ \\
M7 & $0.3$ & $0.7$ & {\bf 0.80} & $1.0$ & $-$ & $-$ & $11.66$ & $11.62$ & $127$ & $0.64$ &
 $0.38$ & $0.64$ & $-0.55$ & $42.3$ & $ 1.1$ & $+$ \\
M8 & $0.3$ & $0.7$ & {\bf 0.70} & $1.0$ & $-$ & $-$ & $12.40$ & $11.49$ & $120$ & $0.46$ &
 $0.57$ & $0.61$ & $-0.61$ & $45.7$ &  $3.5$ & $+$ \\
M9 & $0.3$ & $0.7$ & $0.90$ & {\bf 0.8} & $-$ & $-$ & $11.32$ & $11.57$ & $110$ & $0.70$ &
 $0.36$ & $0.65$ & $-0.55$ & $44.1$ & $28.2$ & $+$ \\
M10& $0.3$ & $0.7$ & $0.90$ & {\bf 1.2} & $-$ & $-$ & $11.20$ & $11.88$ & $160$ & $0.84$ &
 $0.28$ & $0.66$ & $-0.46$ & $41.2$ & $ 3.3$ & $+$ \\
W1 & $0.3$ & $0.7$ & $0.90$ & $1.0$ & {\bf 0.1} & $0.8$ & $10.93$ & $11.64$ & $118$ & $0.6
4$ & $0.30$ & $0.66$ & $-0.52$ & $40.7$ & $ 8.4$ & $+$ \\
W2 & $0.3$ & $0.7$ & $0.90$ & $1.0$ & {\bf 0.2} & $0.5$ & $10.00$ & $11.37$ & $ 70$ & $0.2
5$ & $0.29$ & $0.68$ & $-0.55$ & $48.4$ & $10.2$ & $+/-$ \\
    \hline
\end{tabular}
\medskip

Column~(1) lists the  ID by which we refer to each  model in the text.
Models  whose ID starts  with M  (W) correspond  to CDM  (WDM) models.
Columns~(2) to~(5) indicate the  parameters of the cosmological model,
where  the values  modified with  respect to  the fiducial  model (the
$\Lambda$CDM concordance model) are  in bold face. Columns~(6) and~(7)
indicate  the free  streaming scale  (in $h^{-1}  \Mpc$) and  the dark
matter  particle mass (in  keV), respectively  (for WDM  models only).
Columns~(8) to~(17) are the same as columns~(5) to~(14) in Table~1.

\end{minipage}
\end{table*}

\subsection{The Tully-Fisher zero-point}
\label{sec:TFzero}

We have  argued that  the $M/L_c$  of our fiducial  model are  at best
marginally consistent  with the (zero-point  of) the TF  relation.  We
computed  our TF  constraints from  the data  of TP00,  for  which the
absolute magnitudes have been determined using distances obtained with
the Cepheid period-luminosity relationship.  The zero-point of this TF
therefore  intrinsically contains an  effective Hubble  parameter.  In
order to  be able to  make a comparison  with our models and  with the
data  on the  LF and  the galaxy-galaxy  clustering, we  converted the
absolute magnitudes  of the  TP00 data  to $M_{b_j} -  5 {\rm  log} h$
adopting  $H_0=70 \kmsmpc$.   This  particular choice  for the  Hubble
constant is motivated by (i) the  value $H_0 = 72 \pm 8 \kmsmpc$ (68\%
confidence)  obtained by the  HST Key  Project (Freedman  \etal 2001),
(ii) the value $H_0 = 77  \pm 8 \kmsmpc$ (95\% confidence) obtained by
TP00 from the same TF data used  here, and (iii) the fact that we have
adopted $h=0.7$ in all our models.

However, some uncertainty   regarding $H_0$  remains, and  this   will
directly   influence the  normalization  of  the  TF zero-point.   For
example, determinations of  $H_0$ based on time-delay measurements  of
gravitational lenses  usually yield  values for the   Hubble parameter
that are significantly smaller (e.g. Kochanek 2002; but compare Hjorth
\etal 2002).  Kochanek  (2002) obtains  $H_0 = 48^{+7}_{-4}   \kmsmpc$
(95\%  confidence)  under the   assumption that the  lenses have  flat
rotation   curves.   Furthermore,  Saha  \etal  (2001),  using Cepheid
calibrations  for a sample of  galaxies with Type Ia supernovae, finds
$H_0 = 59 \pm 6 \kmsmpc$.

It is  straightforward to correct our  TF constraints  for an error in
$H_0$. The  mass-to-light  ratios obtained from  equation~(\ref{mlTF})
scale as
\begin{equation}
\label{hscale}
{\rm log}\left({M \over L_c}\right)_h = {\rm log}\left({M \over
L_c}\right)_{0.7} - 2 \, {\rm log}\left({h \over 0.7}\right)
\end{equation}
Thus, for $h < 0.7$ the mass-to-light ratios obtained from the TP00 TF
relation become larger.  In other words,  if for a given halo mass $M$
and our assumption that $h=0.7$ the $M/L_c$ of a model implies a value
$(V_{\rm max}/V_{\rm  vir})_{0.7}$, then for $h \ne  0.7$ the implied
ratio is
\begin{equation}
\label{vratscale}
\left( {V_{\rm  max} \over V_{\rm  vir}} \right)_{h}  = 
\left( {V_{\rm  max} \over V_{\rm  vir}} \right)_{0.7} \times 
\left( {h \over 0.7} \right)^{-0.75} 
\end{equation}
where we have used $b_{\rm TF} = 2.67$ (see Section~\ref{sec:tf}). For
example, in our  fiducial model (M1) the values  of $M/L_c$ imply that
$V_{\rm max}/V_{\rm  vir} \lta 1.4$,  which is hard to  reconcile with
the  typical circular  velocity curves  of CDM  haloes.   If, however,
$h=0.5$ then  we obtain  $V_{\rm max}/V_{\rm vir}  \lta 1.8$,  in much
better agreement with CDM predictions.
\begin{figure*}
\centerline{\psfig{figure=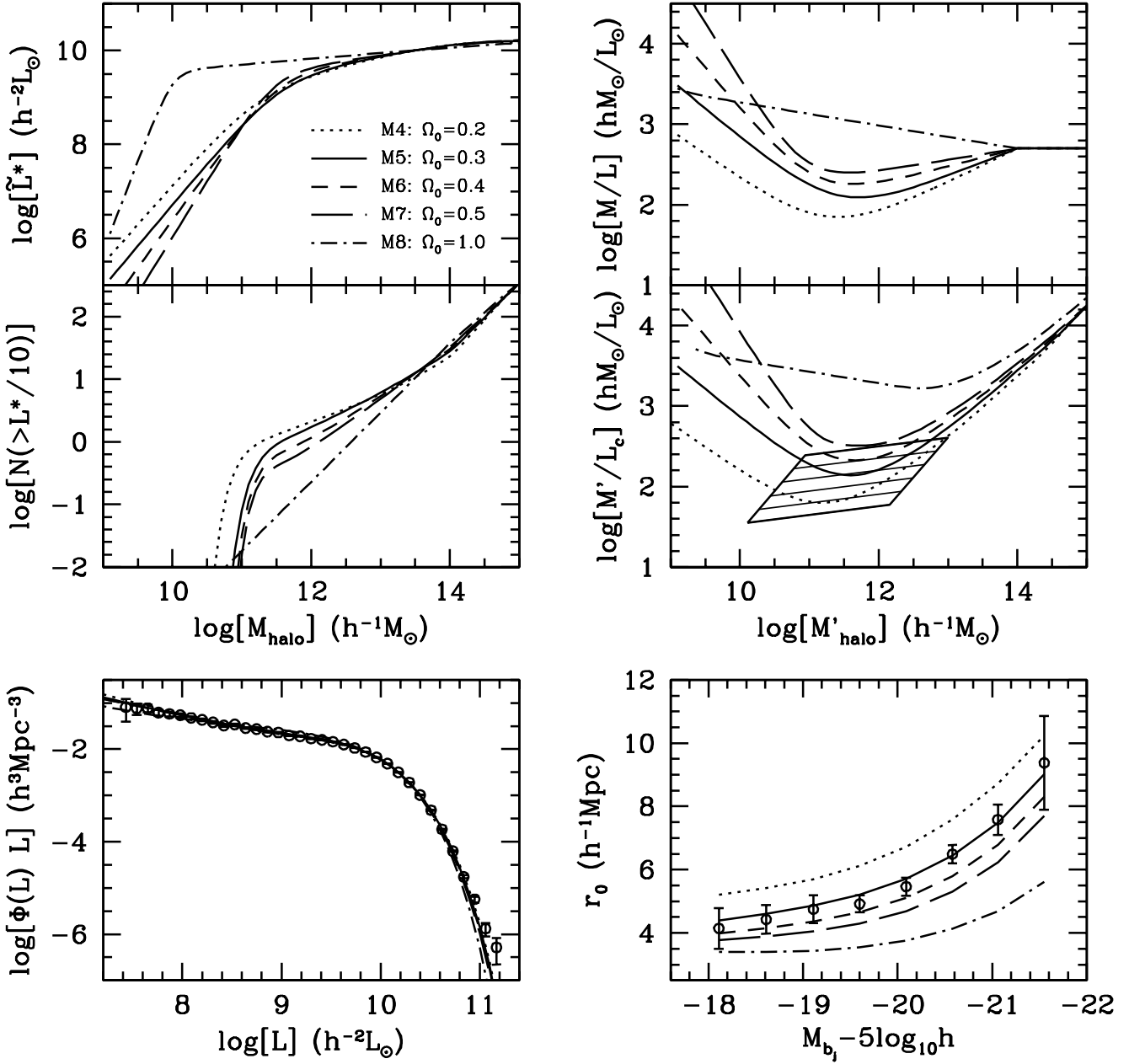,width=\hdsize}}
\caption{Same   as Figure~\ref{fig:concmodel}  except    that here  we
compare results  for a number   of $\Lambda$CDM models  with different
values for $\Omega_0$ as indicated.  All models have $\Omega_{\Lambda}
=   1.0   -  \Omega_0$, normalization   $\sigma_8\Omega_0^{0.52}=0.46$
(Hoekstra \etal 2002), and shape  parameter $\Gamma = \Omega_0 h$ with
$h=0.7$ (see  Table~2). Note that the  mass-to-light ratios (upper two
right   panels)   are  plotted   as    function    of $M'   =     M \,
(\Omega_0/0.3)^{b_{\rm TF}/6}$, and that the middle panel on the right
plots ${\rm log}[M'/L_c]$ rather than  ${\rm log}[M/L_c]$.  This takes
care  of the fact that  the TF  constraints  depend on $\Omega_0$ (see
equation~[\ref{mlTF}]) and allows all models to be plotted in the same
Figure. Note also how model with $\Omega_0 > 0.35$ yield mass-to-light
ratios that  are too high, while models  with $\Omega_0 < 0.35$ result
in correlation lengths that are too large.}
\label{fig:omega}
\end{figure*}
\begin{figure*}
\centerline{\psfig{figure=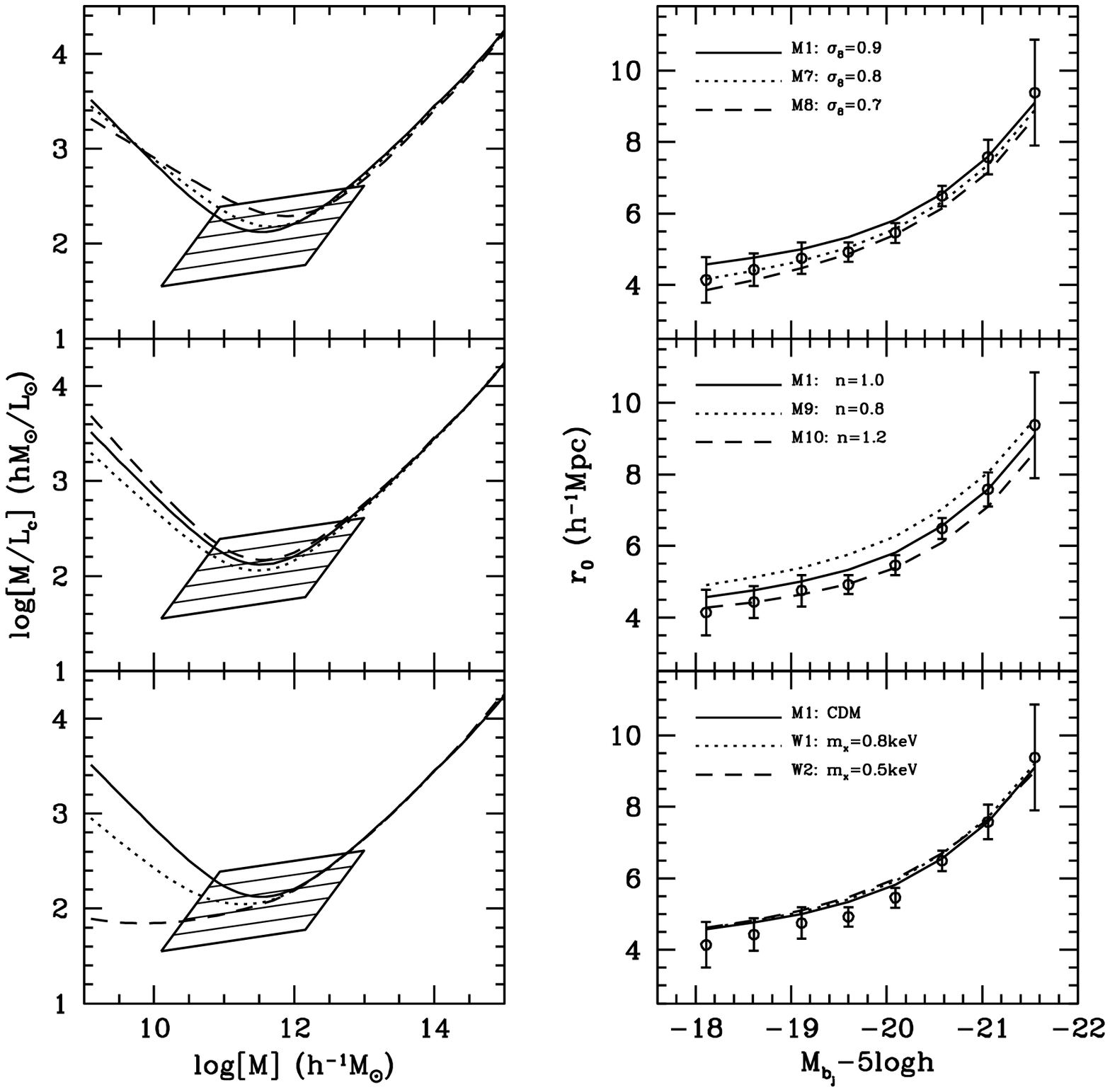,width=\hdsize}}
\caption{Same as Figure~\ref{fig:varmodel} except that here we compare
models in which  we vary $\sigma_8$ (upper  panels), the slope  of the
initial power spectrum $n$ (middle panels), and the nature of the dark
matter (lower  panels) with respect to the  fiducial  model. All other
cosmological parameters  are  kept the same  as  for  the $\Lambda$CDM
concordance cosmology.}
\label{fig:cosmo}
\end{figure*}

\section{Cosmological constraints}
\label{sec:cosmology}

So  far  we  have   only  focussed  on  the  $\Lambda$CDM  concordance
cosmology.  In the present section,  we turn our attention to a number
of  alternative  cosmologies to  (i)  investigate  whether the  method
presented here  can be used  to constrain cosmological  parameters and
(ii) to see whether we can  find a set of cosmological parameters that
improves  the  fit to  $\Phi(L)$  and  $r_0(L)$  with respect  to  the
concordance model  while simultaneously yielding  mass-to-light ratios
that are in better agreement with the TF constraints.

In  a  recent study, Zheng    \etal (2002)  have  shown that  distinct
cosmological models produce  distinct halo  populations.  Based  on  a
detailed investigation  of   halo  populations in  a   wide   range of
cosmologies, Zheng   \etal   argued  that a combination    of   galaxy
clustering  and statistics that are  sensitive to the virial masses of
dark matter haloes should  be  able to  impose strong constraints   on
cosmological    models  without  having  to  rely   on  {\it a priori}
assumptions  regarding galaxy  formation.  Indeed,  recently Guzik  \&
Seljak (2002) and Yang   \etal   (2002) used the galaxy-galaxy    weak
lensing results of the SDSS (McKay \etal 2001) to constrain the virial
masses of   galaxies  brighter than  $L^{*}$    (see also McKay  \etal
2002). Linking  these results to  the abundance of dark matter haloes,
Seljak (2002) was able   to place useful constraints on   cosmological
parameters. This suggests that our combination of constraints from the
luminosity-dependent clustering strengths and the mass-to-light ratios
implied by the TF relation, should  impose constraints on cosmological
parameters.  As we show below, this is indeed the case.

\subsection{The matter density}
\label{sec:omega}

We consider  cosmologies in  which the matter  density is  varied with
respect to the  concordance value $\Omega_0=0.3$ (models M2  - M6; see
Table~2).  All  models considered here have $\Omega_{\Lambda}  = 1.0 -
\Omega_0$,  to keep  agreement  with the  cosmic microwave  background
constraints (e.g.,  Melchiorri \etal  2000; de Bernardis  \etal 2002),
and we adopt  $\Gamma = \Omega_0 h$ throughout.   In addition, when we
modify  $\Omega_0$ we  also  change the  normalization $\sigma_8$,  in
order to  satisfy the  weak-lensing constraints obtained  by Hoekstra,
Yee \& Gladders (2002):
\begin{equation}
\label{sigma8Omega0}
\sigma_8\Omega_0^{0.52}=0.46\,.
\end{equation}
This allows us to investigate  whether our modeling of the conditional
LF  can break the  degeneracy between  $\sigma_8$ and  $\Omega_0$ that
results from weak lensing and cluster abundance measurements.

Comparing   models with  $\Omega_0 \ne  0.3$   with  data is  somewhat
complicated  by  the fact that obtaining   $\Phi(L)$ and $r_0(L)$ from
observations  is cosmology  dependent.   The observational constraints
used   so   far     were  obtained     assuming  $\Omega_0=0.3$    and
$\Omega_{\Lambda}=0.7$,  consistent with  the cosmological  parameters
considered up to  this  point.  For   other values  of $\Omega_0$  and
$\Omega_{\Lambda}$, however,   the observational constraints  from the
2dFGRS will be different.  For the LF,  S.  Cole and P. Norberg kindly
provided  $\Phi(L)$  obtained  from   the  2dFGRS for   the additional
cosmologies   considered   here  (private  communications).  For   the
$r_0(L)$,    however,  only  measurements   for   $\Omega_0=0.3$   and
$\Omega_{\Lambda}=0.7$ are available to us (from Norberg \etal 2001a).
We therefore    ignore  the cosmology    dependence  of  the  observed
correlation lengths in our  analyses.  Since the $\Omega_0$ dependence
is small compared to the statistical errors on $r_0(L)$, this does not
significantly affect our   results. Finally, the TF constraints   also
depend  on $\Omega_0$  (see  equation~[\ref{mlTF}]).   This effect has
been   taken   into account     in    the  middle   right     panel of
Figure~\ref{fig:omega}  by  rescaling $M/L_c$  and  $M$ accordingly so
that  we  can plot the  curves  for different  $\Omega_0$ in  the same
figure.

Figure~\ref{fig:omega} presents results  for the best  fit models with
$\Omega_0=0.2$, $0.3$, $0.4$, $0.5$  and $1.0$.  Models with $\Omega_0
\lta 0.3$ predict values of $r_0(L)$ that are  too large, while models
with $\Omega_0  \gta 0.3$ predict too  high values of $M/L_c$ to match
the observed TF constraints.  Apparently, {\it the technique used here
is  extremely effective in breaking  the degeneracy between $\Omega_0$
and $\sigma_8$} that  is present  in  current measurements of   cosmic
shear and/or  the abundance of  rich galaxy clusters.  Remarkably, the
best results are   obtained for $\Omega_0$  close to  the  concordance
value    of $\Omega_0=0.3$, in good     agreement with the constraints
obtained from high  redshift supernova Ia  searches (e.g., Riess \etal
1998; Perlmutter \etal 1999) and from  observations of the temperature
anisotropies in the  cosmic microwave background  (e.g., Sievers \etal
2002).

For an Einstein-de Sitter cosmology, we find mass-to-light ratios that
are  more  than  an order  of  magnitude  too  high  to match  the  TF
constraints. This is in good  agreement with previous studies based on
a variety of different methods  (e.g., White \& Frenk 1991; Cole 1991;
Kauffmann \etal 1993; Steinmetz  \& Navarro 1999; Elizondo \etal 1999;
van den Bosch 2000).

\subsection{The normalization of the power spectrum}
\label{sec:normalization}

In the previous section we  have shown that models with $\Omega_0=0.3$
yield  the best  results, once  $\sigma_8$ is  kept normalized  to the
constraints obtained  by Hoekstra \etal (2002).   For the $\Lambda$CDM
concordance model this implies $\sigma_8=0.86$.  This normalization is
consistent  with  constraints  from  the observed  abundance  of  rich
clusters of galaxies (e.g. White,  Efstathiou \& Frenk 1993; Eke, Cole
\&  Frenk 1996;  Bahcall  \& Fan  1998;  Viana \&  Liddle 1996,  1999;
Pierpaoli, Scott  \& White 2001)  and from weak  gravitational lensing
studies (van  Waerbeke \etal 2002;  Bacon \etal 2002;  Refregier \etal
2002).  However, a number of  recent studies have suggested a somewhat
lower  normalization, of the  order of  $\sigma_8=0.7$ -  $0.8$ (e.g.,
Carlberg  \etal  1997; Viana,  Nichol  \&  Liddle  2002; Seljak  2001;
Reiprich \& B\"{o}rhringer 2002).  We therefore investigate the impact
of  such small  changes  in  $\sigma_8$, while  keeping  the value  of
$\Omega_0$ fixed at $0.3$.

Changes in $\sigma_8$ mainly affects the clustering strength, which is
proportional to   $b^2 \sigma_8^2$.  The   bias $b(M)$  increases with
decreasing $\sigma(M)$   (Mo \&  White  1996),   and since  $\sigma(M)
\propto \sigma_8$, lowering the  value of $\sigma_8$ causes haloes  of
given mass to  be more strongly   biased.  However, since on  galactic
scales the increase of  $b$  with  $1/\sigma(M)$  is less  rapid  than
linear, the clustering strength decreases  when $\sigma_8$ is lowered.
The upper panels of Figure~\ref{fig:cosmo}  compare our fiducial model
M1 ($\sigma_8=0.9$) with      models  M7  ($\sigma_8=0.8$) and      M8
($\sigma_8=0.7$).   Indeed,  lower   values of   $\sigma_8$  result in
smaller correlation lengths.    In fact, for $\sigma_8=0.8$  the model
yields an  excellent fit to the  observed $r_0(L)$.  In addition, from
the values  of $\chi^2({\Phi})$ it  is clear that these modest changes
of $\sigma_8$ do not significantly influence the quality of the fit to
the LF.  However, the $M/L$ ratios on the  mass scale probed by the TF
relation become slightly higher,  therewith only worsening the problem
of reproducing the correct  TF zero-point.  Thus,  lowering $\sigma_8$
by about 10  percent with respect  to the concordance value brings the
luminosity dependence of the galaxy  clustering strengths in excellent
agreement  with  observations, but     an alternative  or   additional
modification is   required  to bring    the mass-to-light  ratios   in
agreement with the TF zero-point.

\subsection{Tilted CDM models}
\label{sec:tilt}

So far we have only considered models with a Harrison-Zel'dovich power
spectrum ($n=1$). However, inflationary  models allow small deviations
of $n$ from  unity, and we therefore  investigate the effect  of small
modifications of $n$.
 
The  middle  panels of  Figure~\ref{fig:cosmo}  plot  the results  for
models M9  ($n=0.8$) and  M10 ($n=1.2$).  In  both models we  keep the
normalization  of the power  spectrum fixed  at $\sigma_8=0.9$  and we
adopt the cosmological parameters  of the concordance model.  Lowering
$n$ has the  effect of suppressing the linear  power spectrum on small
scales, and so one might hope to reduce the number density of galactic
haloes.   We  find  this  effect   to  be  quite  small,  and  so  the
mass-to-light  ratios required  to match  the luminosity  function are
quite  similar  to  those  obtained  for the  concordance  model  with
$n=1$. The biggest effect of changing $n$ is to change the correlation
length of the mass, for  example, the mass correlation length is about
$5.1\mpch$ for  $n=0.8$, and about $4.7\mpch$ for  $n=1.2$.  Since the
bias factor for galaxies increases with decreasing $n$ (because of the
decrease in $\sigma (M)$  on galaxy scales), the predicted correlation
lengths for  galaxies increases with decreasing $n$,  as shown clearly
in the  middle right panel. Thus,  an increase of the  power index $n$
has a similar effect as lowering $\sigma_8$; it brings the correlation
lengths in  better agreement with the data,  while slightly increasing
the mass-to-light ratios on the scale of galaxies.

\subsection{Warm Dark Matter}
\label{sec:WDM}

In recent years  there have been various claims  that the CDM paradigm
is in conflict with observations  on small scale.  First, CDM predicts
too many low mass haloes which leads to problems with the abundance of
satellite  galaxies around  a typical  Milky Way  sized  galaxy (e.g.,
Moore \etal 1999; Klypin \etal 1999; but see Stoehr \etal 2002) and to
problems with the  formation of disk galaxies (White  \& Navarro 1993;
Navarro  \& Steinmetz  1999,  2000; Mo  \&  Mao 2000;  van den  Bosch,
Burkert  \& Swaters  2001).  Second,  CDM models  predict  dark matter
haloes  with steeply cusped  density distributions,  inconsistent with
the observed rotation curves of  some dwarf and low surface brightness
galaxies (e.g., Blais-Ouellette, Amram \& Carignan 2001; de Blok \etal
2001; but see van den Bosch \& Swaters 2001).

In  order to (partially) alleviate these  problems, Warm  Dark Matter
(WDM)  has been  suggested  as an  alternative  dark matter  candidate
(e.g.,  Sommer-Larsen  \&   Dolgov  2001;  Col\'{i}n,  Avila-Reese  \&
Valenzuela 2000; Bode, Ostriker \& Turok 2000). In WDM models the dark
matter particles have a  non-negligible initial velocity, which causes
the  damping  of   small-scale  fluctuations  due  to  free-streaming.
Consequently  the power-spectrum of  WDM is  modified with  respect to
that of CDM,  and can be expressed in terms of  the CDM power spectrum
$P_{\rm CDM}(k)$ according to
\begin{equation}
\label{pkWDM}
P_{\rm WDM}(k) = T^2_{\rm WDM}(k) \, P_{\rm CDM}(k)
\end{equation}
where the WDM transfer function is approximated by
\begin{equation}
\label{tkWDM}
T_{\rm WDM}(k) = {\rm exp}\left[ -{k R_{f,{\rm WDM}} \over 2} -
{(k R_{f,{\rm WDM}})^2 \over 2}\right]
\end{equation}
(Bardeen  \etal  1986; Sommer-Larsen  \&  Dolgov  2001). The  comoving
free-streaming scale $R_{f,{\rm WDM}}$  is related to a free-streaming
mass $M_{f,{\rm WDM}}$ according to
\begin{equation}
\label{MfWDM}
M_{f,{\rm WDM}} = 3.7 \times 10^{14} \, h^{-1} \, \Msun \, \Omega_{\rm WDM} \,
\left({R_{f,{\rm WDM}} \over h^{-1} \Mpc}\right)^{3}
\end{equation}
which  in turn  is related  to  the mass  $m_X$ of  the WDM  particles
as
\begin{equation}
\label{mWDM}
m_X = 2.4 \, h^{5/4} \, \Omega^{1/2}_{\rm WDM} \,
\left({M_{f,{\rm WDM}} \over 10^{11} h^{-1} \Msun}\right)^{-1/4} \,
{\rm keV}\,.
\end{equation}

The lower two panels of Figure~\ref{fig:cosmo} compare the results for
our fiducial CDM model M1 with  two WDM models, W1 and W2, that differ
only in the  mass of the WDM particles (see  Table~2).  Except for the
properties of the dark matter, these models have the same cosmological
parameters  as the $\Lambda$CDM  concordance model.   Decreasing $m_X$
causes a decrease in the abundance of small mass haloes. Consequently,
in  order to  preserve the  abundance of  (faint) galaxies,  each (low
mass) halo needs to harbor more (or brighter) galaxies.  This causes a
decrease  in the  mass-to-light ratios.   Furthermore, WDM  haloes are
expected to be somewhat  less concentrated than their CDM counterparts
(e.g., Bode, Ostriker \& Turok 2000), favoring lower values of $V_{\rm
max}/V_{\rm  vir}$.  Both  of these  two  effects bring  the model  in
better  agreement with  the TF  constraint.   On the  other hand,  the
change from  CDM to WDM does  not significantly change the  fit to the
luminosity function  and correlation length.   It thus seems  that WDM
models with  $m_X \lta  0.8$ are consistent  with all  data considered
here.  However, such low WDM  particle masses may be hard to reconcile
with  the  observed  opacity  distribution of  the  Ly$\alpha$  forest
(Narayanan  \etal  2000),  with  the phase-space  densities  of  dwarf
galaxies (Dalcanton \& Hogan 2001), and with re-ionization constraints
(Barkana, Haiman \& Ostriker 2001). It thus remains to be seen to what
extent  WDM may  be considered  a successful  alternative for  the CDM
paradigm.

\section{Comparison with semi-analytical models for galaxy formation}
\label{sec:sams}

The problem  of simultaneously  reproducing the galaxy  LF and  the TF
zero-point has been noticed before in studies based on semi-analytical
models for the formation of galaxies. Initially these studies focussed
on Einstein-de Sitter cosmologies,  where the discrepancy was found to
be very large  (e.g., Kauffmann \etal 1993; Cole  \etal 1994). This is
in  excellent   agreement  with  our  results,   which  indicate  that
cosmologies with $\Omega_0=1$ imply mass-to-light ratios that are more
than an order of magnitude too large. Better results were obtained for
cosmologies with lower values of $\Omega_0$, but the overall agreement
remained unsatisfactory (Heyl \etal 1995).

Further improvement was achieved  by including dust obscuration in the
modeling (Somerville  \& Primack 1999; Cole \etal  2000; Benson \etal
2000).  As  long as $V_{\rm max}/V_{\rm  vir} = 1$  was adopted, these
models found good agreement with  the TF relation.  However, using the
more realistic values  in accord with the NFW  profiles of CDM haloes,
the  semi-analytical models  predict too  high rotation  velocities at
given luminosity.   Similar results were  obtained by studies  that do
not attempt  to simultaneously reproduce the galaxy  LF, but normalize
the  mass-to-light ratios  of galaxy-sized  haloes by  adopting baryon
fractions  motivated  by  nucleosynthesis  constraints  (Steinmetz  \&
Navarro 1999; Mo \& Mao 2000; Navarro \& Steinmetz 2000; van den Bosch
2000).  Once  again, our conclusions  are in excellent  agreement with
these results.

In   general,   the    semi-analytical   models   use   the   extended
Press-Schechter formalism  (Bond \etal 1991; Bower  1991; Kauffmann \&
White 1991; Lacey \& Cole  1993; Somerville \& Kolatt 1999) to compute
Monte-Carlo  realizations of  the  halo merger  histories. Since  this
method yields  no information about  the spatial distribution  of dark
matter  haloes,  the clustering  properties  of  galaxies  can not  be
addressed. Various studies  have therefore applied the semi-analytical
models  directly   to  dark   matter  haloes  grown   in  cosmological
simulations (Kauffmann  \etal 1997; 1999; Diaferio  \etal 1999; Benson
\etal 2000, 2001, Springel \etal 2001; Mathis \etal 2001) and computed
correlation functions of the model galaxies.

Benson \etal    (2000)  and  Kauffmann \etal  (1999)    presented  the
mass-to-light ratios  as  function of virial  mass   for their models,
which    can be  directly compared  to    the $\langle M/L \rangle(M)$
obtained  from our  conditional LF. Benson   \etal (2000) find a clear
minimum of  $\sim 125 h  \MLsun$ at $M  \approx 10^{12} h^{-1} \Msun$,
while for  $M \gta 10^{14}   h^{-1} \Msun$ their mass-to-light  ratios
converge  to a value of  $\sim 450 h \MLsun$.   The agreement with our
fiducial model (see upper  right panel of  Figure~\ref{fig:concmodel})
is impressive, suggesting that their  semi-analytical models result in
a conditional LF not  too different from the  one derived here using a
completely different approach.  This  is further supported by the fact
that their models  yield a  luminosity  dependence of  the correlation
length that is in reasonable agreement with our fiducial model.

This  suggests that  the  technique introduced  here,  in addition  to
constraining cosmological parameters, can be used to place constraints
on galaxy formation.  For  example, Kauffmann \etal (1999) showed that
different  treatments of  star formation  and feedback  have  a strong
influence  both  on the  slope  and  the  amplitude of  the  two-point
correlation function  of galaxies.  Benson \etal (2000),  on the other
hand, concluded that the  predictions for the correlation function are
robust to changes  in the semi-analytical model parameters  as long as
the  models match  the bright  end  of the  luminosity function.   Our
results, however,  clearly indicate  that different models,  which fit
the LF  equally well, can have quite  different clustering properties.
For example, models  M1A to M1F all match the  LF roughly equally well
(with  the possible exception  of M1B)  but yield  correlation lengths
that  differ quite  strongly.   This suggests  that,  as concluded  by
Kauffmann \etal (1999), the physics of galaxy formation (which set the
conditional LF)  has a strong  effect on the clustering  properties of
galaxies.
\begin{figure*}
\centerline{\psfig{figure=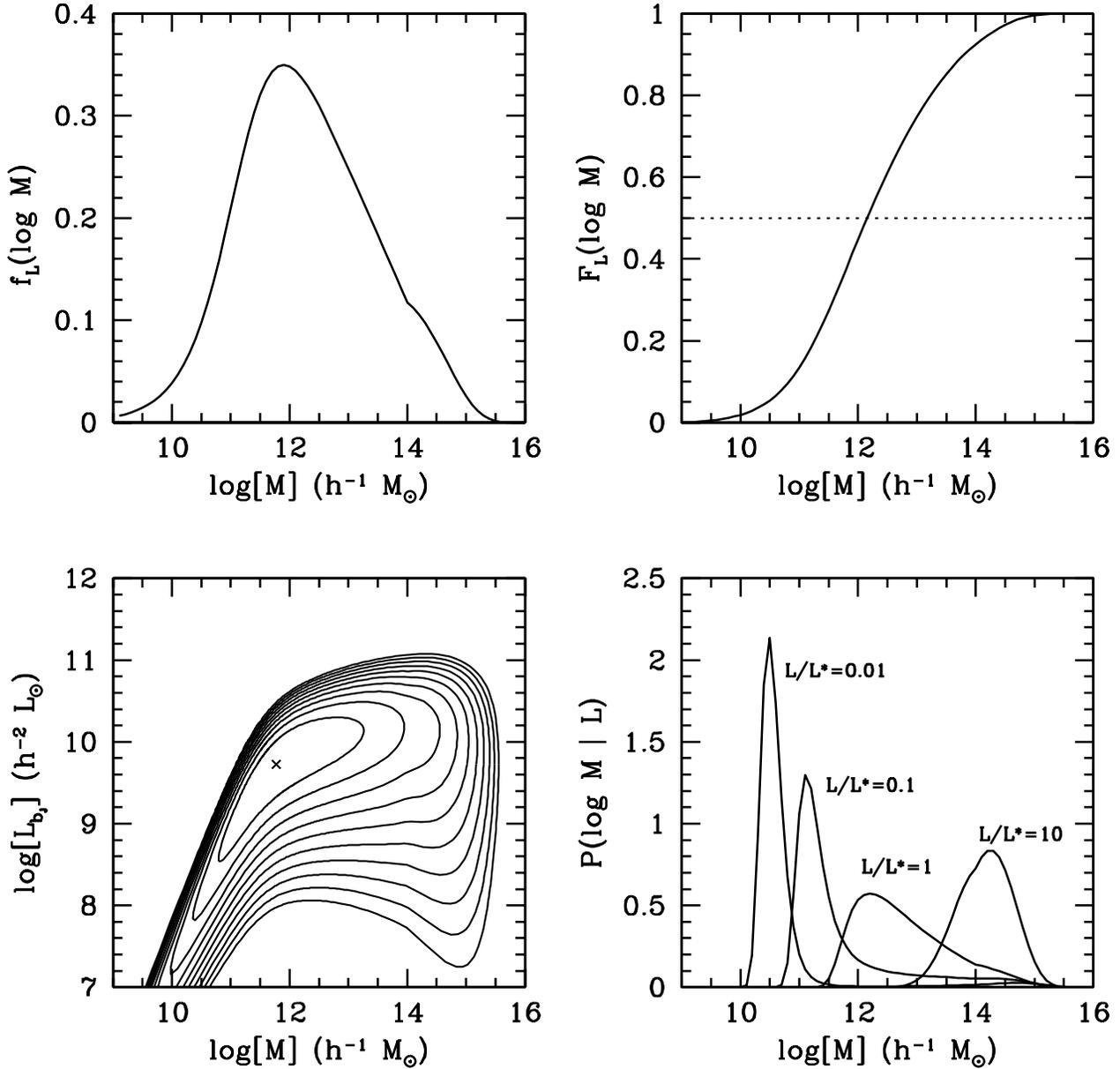,width=0.95\hdsize}}
\caption{Statistics regarding the distribution of blue light at $z=0$.
The  upper left  panel plots  the fraction  of blue  light  emitted by
galaxies inside haloes with virial masses  in the range $M$, $M + {\rm
d}M$. The corresponding cumulative  distribution is shown in the upper
right panel,  indicating that $\sim 50$  percent of all  blue light is
produced inside haloes  with $M < 2 \times  10^{12} h^{-1} \Msun$. The
contour  plot  in  the  lower  left panel  corresponds  to  the  joint
probability  distribution $P(L,M)$  (equation~[\ref{jointprob}]).  The
cross marks the maximum of $P(L,M)$, and the $n$th contour level is at
$2^{-n}$ times the maximum.  Note that galaxies with $L \simeq 10^{10}
h^{-2} \Lsun$  living inside  haloes with $M  \simeq 1  \times 10^{12}
h^{-1} \Msun$  are the most efficient  producers of blue  light in the
local Universe.  Finally, the bottom right panel plots the conditional
probability distributions $P(M  \vert L)$ (equation~[\ref{probM}]) for
four different luminosities as indicated.}
\label{fig:light}
\end{figure*}

The above discussion clearly  indicates the potential strength  of the
technique introduced here.   Note that our conclusions are independent
of the uncertainties present in  current theories of galaxy formation.
In particular, no uncertain assumptions have  to be made regarding the
efficiencies of star formation and feedback,  or regarding the effects
of  merging and  dust  obscuration.  This makes  the results presented
here more robust.  An additional advantage is the computational costs;
only a few minutes are required to find the best-fit parameters of the
conditional LF for  a specific cosmological  model. Thus, our approach
can be  used to   probe a large   number of  cosmological models in  a
relatively short time.  On the other  hand, the approach explored here
has  its limitations. It   only  yields a   parameterized form  of the
conditional LF,  and it is unclear  if the best-fit  $\Phi(L \vert M)$
can be understood in terms  of the physical processes governing galaxy
formation.  This may  be  most easily  addressed with  semi-analytical
models, and as such the two approaches are complementary.

\section{The distribution of light in the local Universe}
\label{sec:light}

The  conditional luminosity function  $\Phi(L \vert  M)$ that  we have
attempted to constrain in this paper is a powerful tool in linking the
distribution  of  galaxies   to  that  of  the  dark   matter.  As  an
illustration, we  use the $\Phi(L \vert  M)$ of our  fiducial model to
present  some statistical  properties  of the  distribution of  (blue)
light (of  galaxies) in  the local Universe  (i.e., at $z  \simeq 0$).
Note that these distributions  only give an approximate description of
the true  distribution of  light.  After all,  they correspond  to our
fiducial model, which is not  perfectly consistent with all data.  The
description given here therefore only serves as an illustration of the
statistical distributions that can be obtained once the conditional LF
is known more accurately.

First  we  use $\Phi(L  \vert  M)$  to  define the  joint  probability
distribution
\begin{equation}
\label{jointprob}
P(L,M) \, {\rm d}L \, {\rm d}M = {1 \over \bar{\rho}_L} \, n(M) \,
\Phi(L \vert M) \, L \, {\rm d}L \, {\rm d}M
\end{equation}
which gives the probability that  a $b_j$-band photon that is produced
in the local Universe originates from a galaxy  with luminosity in the
range $L  \pm {\rm d}L/2$ that  resides in a  halo with virial mass in
the   range $M \pm  {\rm d}M/2$.   Here  $\bar{\rho}_L$ is the average
luminosity density, given by
\begin{eqnarray}
\label{lumdens}
\bar{\rho}_L & = & \int_{0}^{\infty} {\rm d}L \, L  
\int_{0}^{\infty} {\rm d}M \, n(M) \, \Phi(L\vert M) \nonumber \\ 
& = & \int_{0}^{\infty} \Phi(L) \, L \, {\rm d}L\,.
\end{eqnarray}
The contours  in the lower  left panel of  Figure~\ref{fig:light} show
the  joint probability  distribution $P(L,M)$  for our  fiducial model
M1. As can be seen, the joint probability distribution peaks at around
$L \simeq L^{*}$  and $M \simeq 1 \times  10^{12} \msunh$, i.e., these
systems are  the most efficient producers  of blue light  in the local
Universe.   Integrating  $P(L,M)$  over  all luminosities  yields  the
fraction  of all  light, $f_L(M)  \, {\rm  d}M$, produced  by galaxies
inside haloes with virial masses in the range $M \pm {\rm d}M/2$:
\begin{equation}
\label{fraclight}
f_L(M) = \int_{0}^{\infty} P(L,M) \, {\rm d}L =
{1 \over \bar{\rho}_L} \, n(M) \cdot \langle L \rangle(M)
\end{equation}
The  upper  left   panel  of  Figure~\ref{fig:light}  plots  $f_L$  as
function of  halo mass  again for the  conditional LF of  our fiducial
model.  Note  that haloes  with $M \sim  10^{12} \msunh$ are  the most
productive in  `producing' light.  This is not  surprising, given that
$M/L(M)$ reaches  a minimum at around  this mass scale.   This is also
consistent with generic models of galaxy formation, which predict that
haloes more  massive than $10^{12} \msunh$ are  inefficient in cooling
their  baryonic gas,  while  less massive  haloes  are inefficient  in
producing  stars because  of  heating from  supernova feedback  (e.g.,
Larson  1974; White \&  Rees 1978;  Dekel \&  Silk 1986).

The upper  right panel of Figure~\ref{fig:light}  plots the cumulative
distribution $F(<M)=\int_{0}^{M} f_L(M) \,  {\rm d}M$, which gives the
fraction of all light produced in haloes less massive than $M$. In our
fiducial model,  about $50$  percent of all  light is  produced inside
haloes with $M  \lta 2 \times 10^{12} \msunh$.   Since this is roughly
the mass scale where one  changes from isolated haloes around galaxies
to haloes that harbor groups  and clusters of galaxies, this indicates
that isolated  galaxies produce  roughly the same  amount of  light as
galaxies  residing  in  clusters  and  groups.   Note  also  that  the
distribution of $f_L(M)$ is quite  narrow.  Galaxies in haloes with $M
\gta 2  \times 10^{14}  \msunh$ and $M  \lta 5 \times  10^{10} \msunh$
each are responsible for only about  5 percent of the total blue light
in the local Universe.

Finally,  we  use  the  conditional  LF  to  compute  the  conditional
probability  distribution $P(M  \vert L)  {\rm d}M$  that a  galaxy of
luminosity $L$  lives inside a halo  with virial mass in  the range $M
\pm {\rm d}M/2$.  Using Bayes' theorem,
\begin{equation}
\label{probM}
P(M \vert L) \, {\rm d}M = {\Phi(L \vert M) \over \Phi(L)} 
\, n(M) \, {\rm d}M\,.
\end{equation}
The lower right panel of Figure~\ref{fig:light} plots this probability
distribution for four  different  luminosities: $L = L^{*}/100$,  $L =
L^{*}/10$,   $L  = L^{*}$,  $L =   10 \, L^{*}$.    Whereas $10 L^{*}$
galaxies  are typically found in  haloes  with $10^{13} h^{-1} \lta  M
\lta 10^{15} h^{-1} \Msun$, galaxies with $L = L^{*} / 100 \sim 10^{8}
h^{-2} \Lsun$ typically reside in haloes of $M \simeq 5 \times 10^{10}
\msunh$.  Only a  very small  fraction  of  these  galaxies ($\lta  5$
percent) are   found  as a member of    a larger group or  cluster  of
galaxies.  It   should be    pointed  out, however,  that   our  model
predictions for $P(M \vert  L)$ are quite  uncertain for galaxies with
$L \lta L^{*}/10$. This owes to the fact  that constraints on $r_0(L)$
only exist for galaxies  with $L >  L^{*}/5$, even though data on  the
luminosity function extends down  to  much fainter luminosities.   If,
for example,  for $L < L^{*}/5$  the correlation lengths increase with
decreasing  luminosity (i.e., opposite to the  trend  for the brighter
galaxies), we would obtain conditional  luminosity functions with many
more faint galaxies in massive haloes than  in the current prediction.
Clearly, observational data  on   the correlation functions  of  faint
galaxies are needed in order to better constrain their $P(M \vert L)$.

\section{Conclusions}
\label{sec:concl}

We have  presented a  novel   technique to link the   distribution  of
galaxies  to that of dark matter  haloes.  This technique extends upon
earlier modeling of halo occupation  number distributions by labelling
the galaxies with luminosities.  In particular, we have introduced the
conditional luminosity  function $\Phi(L  \vert  M) {\rm  d}L$,  which
gives the average number of galaxies with luminosities in the range $L
\pm {\rm d}L/2$ that reside  in a halo  of mass $M$. Starting from the
number density  of CDM haloes, $n(M) {\rm  d}M$, we sought the $\Phi(L
\vert M)$ that   reproduces the 2dFGRS $b_j$-band luminosity  function
$\Phi(L) {\rm d}L$.  Since there are many  different $\Phi(L \vert M)$
that are   consistent  with both  $n(M)$ and   $\Phi(L)$,  we used the
luminosity dependence of  the galaxy correlation lengths obtained from
the 2dFGRS as additional constraints.   This allows the recovery of  a
unique best-fit $\Phi(L  \vert M)$.   To assess  the viability of  the
conditional  LF thus found, we used  $\Phi(L \vert  M)$ to compute the
average mass-to-light ratios   of dark matter  haloes  as  function of
mass,  which we compared  to constraints obtained from the (zero-point
of) the TF relation.

For   the    $\Lambda$CDM   concordance   cosmology   ($\Omega_0=0.3$,
$\Omega_{\Lambda}=0.7$,  $h=0.7$,  $\sigma_8=0.9$),  we  are  able  to
accurately reproduce  the observed LF. In addition,  adopting a simple
model for the  distribution of galaxies inside dark  matter haloes, we
obtain galaxy-galaxy  correlation functions as  function of luminosity
that  are  in  agreement  with  the data  of  Norberg  \etal  (2001a).
However, the values of $\langle  M/L \rangle$ we obtained are slightly
too high to match that implied by the TF relation.
 
We have shown  that the concordance model can  be made consistent with
the  data if  either the  2dFGRS is  only $\sim  70$  percent complete
(which is very  unlikely), or if $\sim 30$ percent  of all dark matter
haloes  have failed  to form  observable galaxies  (they  only contain
`dark' galaxies).   Although feedback  and an ionizing  background can
explain the presence  of such dark galaxies in  low mass haloes, there
is currently  no clear  mechanism that can  prevent star  formation in
more massive haloes,  rendering this option unlikely as  well (but see
Verde \etal 2002).

The mass-to-light ratios of the concordance model  can also be brought
into agreement with the TF constraints if $H_0$ is lowered to $\sim 55
\kmsmpc$.  Although this value is only  marginally consistent with the
recently advocated value of $H_0 = 72 \pm 8 \kmsmpc$ (68\% confidence)
of  the HST Key Program  (Freedman \etal 2001),  such  low value is in
agreement with recent  time-delay  measurements in gravitational  lens
systems  (Kochanek  2002;   but  see     also Hjorth   \etal    2002).
Alternatively, $\langle M/L \rangle$ may be  lowered by adopting warm,
rather than cold dark matter.  In WDM cosmologies the abundance of low
mass haloes  is suppressed,  resulting in lower  mass-to-light ratios;
WDM haloes are  also expected to  be less concentrated than CDM haloes
and so the mass-to-light ratios implied by the observed TF relation is
also  higher.  Both of these  effects lead to better agreement between
the model  predictions  and the TF constraints.   In  order to  have a
significant effect, we find that the WDM particles must have a mass of
the order of $m_X \la 0.8$keV, which is already the lower limit set by
a number of other   observational constraints (Narayanan  \etal  2000;
Dalcanton \& Hogan 2001; Barkana \etal 2001).

In addition  to these modifications of  the  concordance cosmology, we
have also  investigated cosmologies  with different matter  densities.
We have  shown that the technique  presented here  allows one to break
the degeneracy  between   $\Omega_0$ and $\sigma_8$   inherent  in the
cosmological constraints obtained   from cosmic shear and/or   cluster
abundance  measurements.  The best  results are obtained for $\Omega_0
\simeq  0.3$,  in good agreement   with   the high redshift  supernova
results (e.g., Riess \etal 1998; Perlmutter \etal 1999): for $\Omega_0
\gta 0.4$ the mass-to-light ratios of dark matter  haloes are too high
to match the TF zero-point, while cosmologies with $\Omega_0 \lta 0.2$
yield correlation lengths that are too large.

In summary, the modeling of the conditional LF presented here strongly
favors a cosmology with  $\Omega_0 \simeq 0.3$.   However, in order to
bring the models into agreement with the observed zero-point of the TF
relation, either $h$ is  lowered by 15 to 20  percent with  respect to
the concordance values of $0.7$, dark matter is warm rather than cold,
or a large fraction of haloes host no luminous galaxies.

Once a satisfactory  model for the conditional LF  is found, it can be
used to extract useful  statistical information about the distribution
of  light  in the  Universe.   As an   example, we  have  computed the
fractional contribution to  the  total luminosity density by  galaxies
with luminosity $L$ that live in haloes of mass $M$. In particular, we
have shown that 50 percent of all the blue light is produced in haloes
with  $M \lta   2  \times 10^{12} h^{-1}   \Msun$.   In  addition, the
conditional LF can  be used to compute the  probability $P(M \vert  L)
{\rm d}M$ that a galaxy with luminosity $L$ lives  in a halo with mass
in the range $M \pm {\rm d}M/2$.   These probability distributions may
prove useful for  a  statistical interpretation of  the  mass-to-light
ratios inferred from, for example,  galaxy kinematics or gravitational
lensing.

The  technique presented  here can   easily  be  extended  to  compute
higher-order  correlation functions,  as well  as  pair-wise  peculiar
velocity dispersions,  all as function of  galaxy luminosity.   In the
near future  several  large redshift surveys  of  galaxies will become
available, such as the SDSS and the completed 2dFGRS at $z=0$ and DEEP
and VIRMOS at $z=1$.  These  will allow far more stringent constraints
to be placed on cosmological parameters, especially when combined with
the    complementary    constraints  from    weak-lensing  and cluster
abundances.   In addition,  once  luminosity functions and correlation
lengths  are   available in different   photometric passbands  and for
different   morphological classes,  the   technique presented here can
easily be   extended   to simultaneously  model the    distribution of
galaxies as  function of color and/or  morphological type.   This will
not only   help  to constrain  cosmological   parameters, but also  to
provide important insights  into the complicated  processes associated
with galaxy formation.


\section*{Acknowledgements}

We  are grateful to  Shaun Cole  and Peder  Norberg  for providing the
2dFGRS luminosity functions  in  several cosmological models used   in
this  paper.  We  thank Simon  White  for insightful  comments and for
carefully reading   the  paper, and Andreas    Berlind, Peder Norberg,
Rachel Somerville,   David Weinberg,  and Saleem  Zaroubi   for useful
discussions.    XY thanks the   MPG-CAS  student exchange  program for
financial support.


\end{document}